\documentclass[final,twocolumn,11pt]{elsarticle}
\usepackage{lscape}
\journal{Icarus}
\begin{document}
\begin{frontmatter}
\title{Thirty Years of Cometary Spectroscopy from McDonald Observatory} 
\author[UTA]{A. L. Cochran\corref{cor1}}
\ead{anita@barolo.as.utexas.edu}
\author[LZT]{E. S. Barker}
\ead{ebarker40@comcast.net}
\author[NMSU]{C. L. Gray}
\ead{candaceg@nmsu.edu}
\address[UTA]{McDonald Observatory, University of Texas, 1 University Station 
C1402, Austin, TX 78712}
\address[LZT]{LZ Technology, Inc., 1110 NASA Parkway Suite 650, Houston,
TX 77058}
\address[NMSU]{New Mexico State University,
1320 Frenger Street, Las Cruces, NM 88003 }
\cortext[cor1]{Corresponding Author}

\begin{abstract}
We report on the results of a spectroscopic survey of 130 comets that
was conducted at McDonald observatory from 1980 through 2008. Some
of the comets were observed on only one night, while others were observed
repeatedly.  For 20
of these comets, no molecules were detected.  For the remaining 110
comets, some emission from CN, OH, NH, C$_{3}$, C$_{2}$, CH, and NH$_{2}$
molecules were
observed on at least one occasion.  We converted the observed molecular
column densities to production rates using a Haser (1957) model.  
We defined a restricted data set of comets that had at
least 3 nights of observations.  The restricted data set consists
of 59 comets.  
We used ratios of production rates to study the trends in the data.
We find two classes of comets: typical and carbon-chain depleted
comets.  Using a very strict definition of depleted comets, requiring
C$_{2}$ \underline{and} C$_{3}$ to both be depleted, we find 9\%
of our restricted data set comets to be depleted.  Using a more relaxed
definition that requires only C$_{2}$ to be below a threshold (similar
to other researchers), we find 25\% of the comets are depleted.
Two-thirds of the depleted comets are Jupiter Family comets, while
one-third are Long Period comets.  
37\% of the Jupiter Family comets are depleted, while 18.5\% of the Long Period
comets are depleted.
We compare our results with other studies and find good agreement.
\end{abstract}
%{\bf Proposed Running Head}: Comet Observations from McDonald Observatory 

\end{frontmatter}

\section{Introduction}

Since the time that the
first spectra of comets were obtained, it has been remarked that
all cometary spectra are very similar, with the most marked difference being
the amount of continuum present.  This simple observation leads to the question
of whether all comets share the same composition or if there are comets with
fundamentally different compositions.  If differences are seen, then an
additional question of the origin of the differences would be raised: are
differences the result of different formational scenarios or different
evolutionary scenarios?  These questions are not just curiosities.  We
believe that comets are leftovers from the formation of the Solar System, that
they have undergone only small changes since their time of formation, and that
they represent the building blocks of the jovian planets.  Thus, understanding
their chemistry is important for our understanding of the conditions at the
start of the formation of the Solar System.

With these questions in mind, we began a program in 1980 to obtain spectra of
comets using low resolution spectrographs at McDonald Observatory. This program
continued through 2008.  We obtained observations of 132 different comets,
some at more than one apparition.  This paper reports on the observations of
130 of those comets. 29P/Schwassmann-Wachmann~1 is left out because its 
almost-circular, 5\,AU orbit and random outbursts make it an unusual
comet that has
been described in several previous papers (Cochran {\it et al.} 1980, 1982,
1991a, 1991b; Cochran and White 1993). 
\nocite{cobaco80,cocoba82,coco91sw1p1,cocobast91,cowh93}
D/Shoemaker-Levy 9 broke into many pieces prior to impacting Jupiter in 1994. 
Observations of the pieces had to be obtained in a different manner from the
other comets and these are detailed in Cochran {\it et al.} (1994).
\nocite{cosl994}

Some of the comets described in this present paper were already discussed in
prior papers (e.g. Cochran {\it et al.} 1987, 1989, 1992, 2009; Cochran and
Barker 1987).  They are included again for completeness so that all the
observations can be found in the same place.  In addition, all observations
were analyzed in a consistent manner in this paper. Thus, this paper
supersedes its predecessors.
\nocite{coba87gz,cobarast92,cobacari09,cogrba89,co87corr}

This is not the only survey of its kind to be carried out.  A'Hearn et al.
(1995, hereafter AH95) reported on observations of photometry of 85 comets.
Fink (2009, hereafter F09)
presented spectroscopic observations of 92 comets. Langlund-Shula and
Smith (2011, hereafter LS11) observed 26 comets.  We will compare our
results with these studies near the end of this paper.
\nocite{ahetal95,fi09,lasm11}

\section{Observations}

The observations described in this paper were obtained with three different
instruments on two telescopes at McDonald Observatory.  The vast majority of
the observations were obtained with two spectrographs at the Cassegrain focus
of the 2.7m Harlan J. Smith telescope.  A few observations were obtained at
the Cassegrain focus of the 2.1m Otto Struve telescope.

The earliest observations were obtained with the Intensified Dissector Scanner
(IDS) spectrograph on the 2.7m telescope.  This instrument had two $4 \times 4$
arcsec apertures,
separated by 52 arcsec along an east/west line, that were imaged on a
photocathode.  Spectra were obtained by rapidly scanning the complete
wavelength range, which was typically 3200 -- 6000\AA, at 11\AA\ resolution.
A complete spectral scan took 50 sec so all wavelengths were essentially
observed simultaneously. The comet optocenter was alternately imaged through
the ``A'' slit with the ``B'' slit imaging 52 arcsec away in the coma.  Then
the telescope was nodded to image the comet in B and coma in A.  The sequence
generally went ABBA.  Sometimes, the telescope was set so that the comet
optocenter was between the slits or at some other known position and spectra
were obtained without nodding the telescope.  Scans were repeated until
sufficient signal/noise was obtained.

The photocathode made the instrument extremely sensitive.  Thus, we were able
to obtain spectra of comets as faint as V=19.5.  However, it also made it
impossible to observe very bright objects.  Thus, our standard stars were
limited to those that were fainter than V=10 and were taken from the list
of Stone (1977).  In addition, we were not able to obtain observations of
solar analog stars with this instrument (none are faint enough) and had to
use fluxes from the Arvesen {\it et al.} (1969) atlas convolved with the
slit function of the instrument.  The Arvesen {\it et al.} atlas is a
whole-disk integrated high-resolution atlas suitable for our purposes. It is
not the only available atlas and there are known differences in the flux
calibration between the Arvesen {\it et al.} atlas and other atlases, especially
below 4000\,\AA. Thus, the Arvesen {\it et al.} atlas might not be
a perfect match for the Sun.  However, because we color weight the solar
spectrum (see discussion below) before we use it, the most critical
feature for our purposes is that the lines are correct in wavelength
and strength.  The Arvesen
{\it et al.} atlas is quite acceptable for our purposes.
\nocite{st77,argrpe69}

The second instrument to be used was the Large Cassegrain Spectrograph (LCS)
at the 2.7m telescope.  This was a long slit CCD spectrograph with various
gratings.  The bulk of the observations were obtained with a grating set to
cover the spectral range from 3000 -- 5700\AA\ at 7.5\AA\ resolution. 
Occasionally, we tilted the grating further to the red to observe other
features or used a higher resolving power grating to study some bands in more
detail. The slit was $\sim140$\,arcsec long and could be oriented at any
position angle by rotating the instrument on the back of the telescope. Each
pixel on the detector covered a spatial interval of 1.28\,arcsec. The slit
width was variable but was typically set to 2\,arcsec for comet observations
and 10\,arcsec for standard star observations (so that no flux was lost). 
Generally, solar analog stars were observed at least once per run and often
every night.  When not observed for whatever reason, convolution of the slit
function with the Arvesen {\it et al.} (1969) atlas was used for a
solar spectrum.

With the LCS, we would always observe with the optocenter imaged on the chip
at least once every night.  We would then sometimes translate the slit under
very accurate telescope control to extend our coverage into the coma.  Some
nights we also observed at more than one position angle.

The third instrument was the Electronic Spectrograph Number 2 (ES2), used on
the 2.1m telescope.  This instrument is very similar to the LCS and even used
the same detectors. The slit position angle could also be changed by rotating
the instrument on the back of the telescope.  Unlike the LCS, we could not
image additional regions of the coma because the 2.1m telescope cannot be
positioned accurately enough. 

\begin{figure}
\includegraphics[scale=0.35,angle=270]{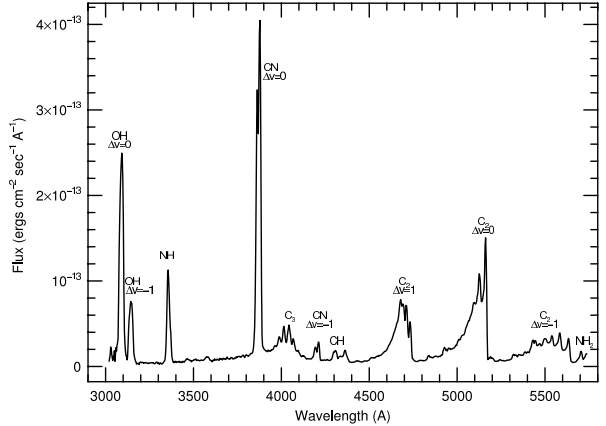}
\caption[fig1]{A spectrum of comet 122P/deVico obtained with
the LCS on 27 September 1995 UT when the comet was at 0.69\,{\sc au} from
the Sun.  The various molecular bands in our bandpass are marked on the
spectrum. }
\label{devico}
\end{figure}

Figure~\ref{devico} shows a spectrum obtained with the LCS of comet 
122P/de~Vico on 27 September 1995 UT. 
This is a very clean spectrum
covering $1.28 \times 2$\,arcsec around the optocenter.  The various
features observed in our bandpass are 
noted on the figure.  These are all molecular emission bands and are composed
of many individual lines that we do not resolve.
This comet had a very low dust-to-gas ratio so there is 
very little continuum to be seen.  The OH, NH and CN features tend to be 
relatively sharp and narrow and are easy to see and measure.  The C$_{3}$ and 
C$_{2}$ bands are broad and can easily be lost in noise when weaker, in
spectra with lower signal/noise or when the continuum due to sunlight
reflecting off the dust is very strong.  The latter is a problem because
the continuum removal can represent a large source of uncertainty and
add noise to the result, even when the signal/noise of the comet
spectrum is very high.  The continuum decreases faster with cometocentric
distance than does the C$_{2}$ or C$_{3}$ band strength, so this is more of
a problem on the optocenter points than further into the coma.
The CH is always very weak and difficult
to measure.  In the vast majority of the spectra reported in this paper,
our bandpass included the features due to CN, C$_{3}$, CH, and C$_{2}$.  We
could never measure OH and NH$_{2}$ simultaneously because, as shown in the
spectrum, the wavelength separation is beyond the bandpass of the instrument
(only part of the
NH$_{2}$ is seen on this spectrum).  In addition, high airmass makes
observation of the OH very difficult (though ironically, the de~Vico spectrum
was obtained when the comet was at 2.5\,airmasses and we were able to observe
the OH in this comet even when it was at 3.8\,airmasses (!) because the comet
was so close to the Sun).

The IDS was used from 1980 through 1989.  We started using the LCS in 1985. 
Thus, there was a period from 1985 -- 1989 when we interchanged which
instrument was used (for reasons that have long since been forgotten!). 
The LCS was used with an RCA $512 \times 512$ pixel CCD from 1985 through 1986.
This detector did not have UV response so was used in a redder setting but
with more limited bandpass because of the smaller number of pixels than
the subsequent CCD. 
From 1988, we used a Texas Instruments $800 \times 800$ pixel CCD that was
UV sensitized.  We also used the Texas Instrument CCD on ES2.  In all cases,
the telescope was tracked at the comet's rate of motion on the sky and we
guided the telescope directly off the slit.  Thus, positional accuracy was
generally quite good.

As would be expected with a survey that spanned so many years, we changed our
observational philosophy during the course of the survey.  When we started, we
wanted to be as inclusive as possible and would observe as many comets to
V$\sim$19.5 as we could.  We therefore sometimes observed comets at heliocentric
distances greater than 2.5\,AU, where they were generally not very active. 
We sometimes also sacrificed signal/noise for the opportunity to observe
another comet. We might only observe a particular comet on a single night. 

As our survey progressed and as we changed instruments, we tended to
concentrate more on targets that were brighter and at smaller heliocentric
distances.  We chose opportunities to observe the same comet repeatedly over
opportunities to observe additional comets.  In addition, comets of special
interest (mission targets, close approaches, etc.) began to dominate the
survey.  Starting in 1995, whenever a comet was bright enough, we would observe
at very high spectral resolution instead of with the instruments described
here.  Those observations are not included in this paper.

\section{Reduction and Analysis}

The routine reduction of the data is very similar for each of the instruments.
First, we removed the bias for the CCDs (the IDS did not have a bias) and flat
fielded the data.  Since the CCD instruments (LCS and ES2) are long slit
instruments, we extracted the stellar spectra using variance weighting
(an estimation of the variance created from a noise model
based on the gain and readnoise parameters and a smooth profile function)
of the slit profile to collapse the stellar spectral image into a 1D spectrum.
The comet spectra for the CCD instruments were left as 2D spectral images.
For all instruments, we observed arc lamps in order to define the
wavelength correspondence of each pixel.  For the IDS, we used a fifth or
sixth order polynomial to fit the dispersion.  For the LCS, we needed to
fit the wavelength in both the spectral and spatial directions.  We used a
fifth order polynomial in the spectral direction and a fourth order
polynomial in the spatial direction.  We would use our spatial fit to
transform the spectral image so that the spectra were aligned with
rows and the spectrum was no longer
curved along the chip.  Finally, the cometary and solar
analog spectra were flux calibrated with the standard star observations
to convert the spectra into flux versus wavelength.

The cometary spectrum consists of an amalgamation of two types of
contributions.  The coma gas yields a molecular band spectrum in
emission, generally from resonance fluorescence.  Sunlight reflecting off
the dust creates a continuum underlying the gas spectrum.  In addition,
there is a third spectrum superimposed on the cometary spectrum and that is
the spectrum of the Earth's atmosphere (the telluric night sky emission spectrum).  

For the purposes of this paper, we are interested in the quantities of the
gases that are imaged through our aperture.  Thus, we need to remove both the
telluric night sky emission spectrum and the solar spectrum. 
The most prominent telluric feature in
our spectrum is the O~($^1$S) feature at 5577\AA.  The telluric spectrum
also has bands in the UV that must be removed.  To remove the telluric
spectrum, we would use either a dedicated sky observation (always
the case for the IDS), obtained well
removed from any comet, or we would use the spectrum at the end of the long
slit (for the LCS and ES2) if a comet was not very extended and there were
no gas features seen in that part of the spectral image.   
We would
weight the sky spectrum by using the strength of the 5577\AA\ band above
the local continuum and then would subtract the sky spectrum from the comet
spectrum.  Note that, for many comets, there is the C$_{2}$ $\Delta
v=-1$ band in the region of the 5577\AA\ line so we needed to define the
continuum with respect
to this band.  Because the 5577\AA\ feature is by far the dominant telluric
signal, slight mis-weighting of the telluric spectrum was not terribly
critical except for the C$_{2}$ $\Delta v=-1$ band.  We never use this band
in our analysis because
there could be uncertainties induced in it by this process.
We see no other strong telluric features (McDonald Observatory does not suffer
from street lighting induced features). There are some weak features
seen, including the Herzberg bands of O$_2$, predominantly below 3600\AA.

Some comets have very strong continuum spectra because their comae contain
a great deal of dust to reflect sunlight; others have very weak continua.
To remove the solar spectrum, it would be ideal to have observations of the
Sun observed with the same instrument.  This is not possible.  The next
best thing would be to observe a star that is a strong match to the
properties of the Sun.  None of the tabulated solar analogs is fainter
than V=10 so we were not able to observe them with the IDS, as noted above.
For the LCS and ES2, we could observe solar analogs but considerations such
as weather or availability meant that we did not always observe one.  The
final resort for a solar spectrum was to use a full disk spectrum of the
Sun.  These are typically observed at high spectral resolution and must be
convolved with a slit function to match our instrumental resolution.  We
adopted the solar spectrum of Arvesen {\it et al.} (1969) for this purpose.

It should be noted that the small particles of dust in the coma result in
the spectrum of the reflected sunlight being reddened when it is reflected
off the dust.  Thus, the solar spectrum does not exactly match the
continuum we wish to remove.  To fix this problem, we measured the flux in
up to 11 continuum bandpasses in both the comet and the solar spectrum and
weighted
the solar spectrum to match the same color.  Use of so many continuum
regions allows us to match any non-linear reddening that exists.
We used a smooth spline function to fit these continuum regions.  The
resultant weighting curve was never a straight line.  This color weighted
spectrum was then removed from the comet spectrum.
This process would alleviate any issues with the UV-blue flux calibration
of the Arvesen {\it et al.} (1969) atlas or with stars of slightly
different colors than the Sun.

\begin{figure}[ht!]
\includegraphics[scale=0.45]{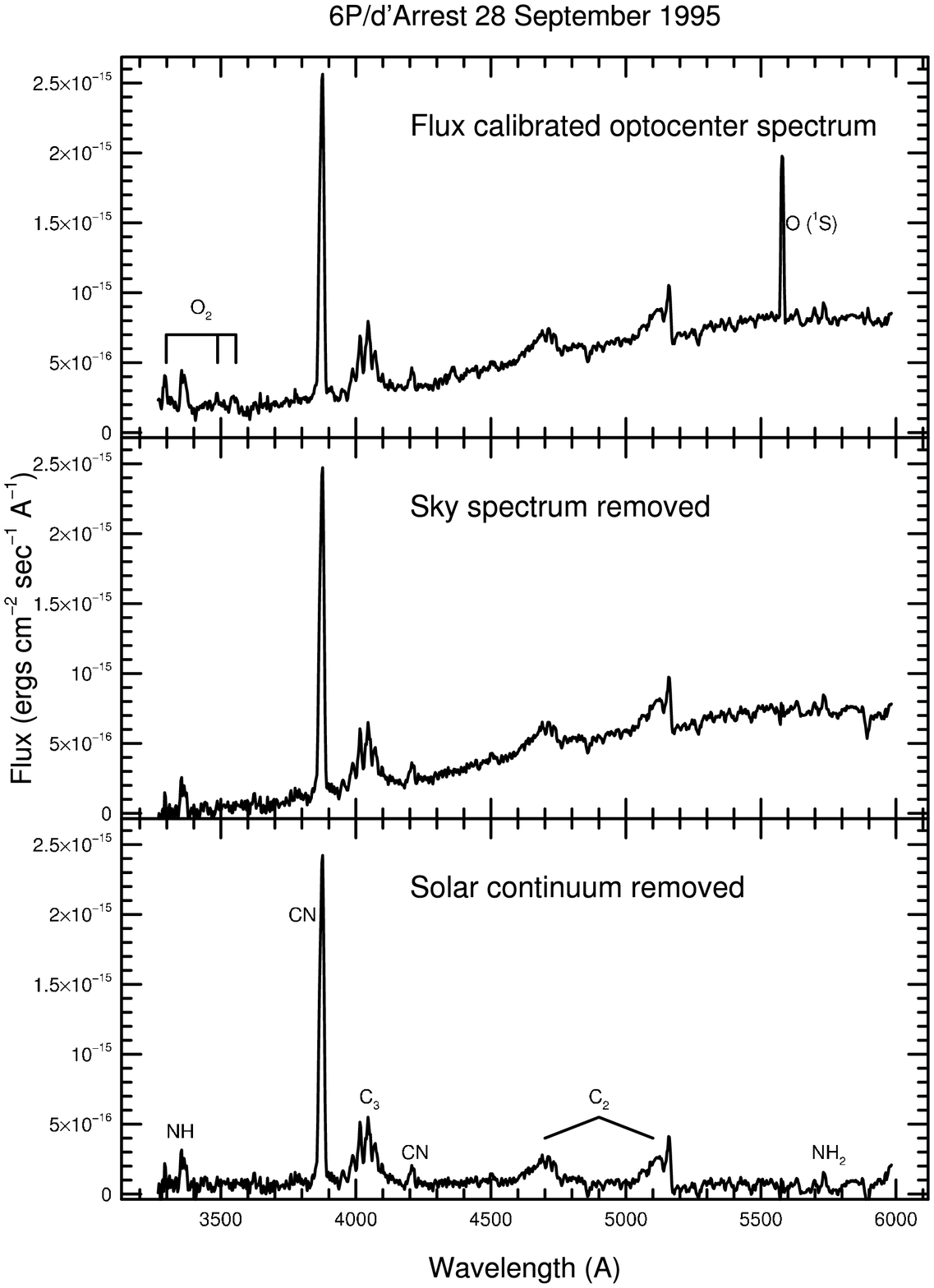}
\caption[fig2]{This figure demonstrates the process of
removal of the telluric night sky emission and solar spectra from the original
spectrum of
comet 6P/d'Arrest.  The top panel shows the flux and wavelength calibrated
spectrum of the comet before removal of either of these
spectra.  Note the strong feature at 5577\AA.
This is the O ($^1$S) band in the telluric spectrum.   The O$_2$
Herzberg bands are also marked. The continuum results
from sunlight reflecting off the dust.  The middle panel shows the same
spectrum but with the telluric spectrum removed using an independent
observation of the sky.  The bottom panel shows that spectrum after the
removal of the solar continuum.  In this case, the spectrum of solar
analog Hyades 64 was removed after weighting the continuum to match the
colors of the star and the dust (see text).}
\label{process}
\end{figure}

In the case of the IDS, each observation would contain spectra of two
positions - often the optocenter and 52 arcsec away.  Each position was
handled separately in the process described above. The off position
still included the coma of the comet so was not used as a source of
a sky spectrum unless the off spectrum was devoid of any emissions.
For the LCS and ES2,
the long slit covers $>$110 pixels spatially.  For high signal/noise
LCS or ES2 spectra, we would treat each spatial pixel as a separate
spectrum for the
telluric and solar removal.  For lower signal/noise observations, we would
sometimes bin together pixels in the spatial direction and then apply the
removal of the telluric and solar spectra (the concept of binning
does not apply to the IDS).  Figure \ref{process} shows an
example of handling the optocenter pixel for comet d'Arrest on 28 September
1995.  
The optocenter spectrum has the strongest continuum.  The continuum
due to the dust weakens faster with spatial distance from the optocenter
than does the gas emission.

Once the telluric and solar spectra were removed, we could now measure the
quantity of gas in each emission band.  First, we fit a continuum to the
region around each of the molecular bands of interest, removed the continuum
and integrated the flux above the continuum.  Though there should be no
continuum left after removal of the telluric and solar spectra,
in reality, the color
weighting is imperfect because there are few true continuum regions for the
weighting.  The fitted continuum at this point provides a consistent level
to integrate above.  

The integrated fluxes are then converted to column densities using the
standard efficiency factors listed in Table~\ref{gfactors}. 
\begin{table*}[ht!]
\caption{Constants for Converting Fluxes to Column Densities} \label{gfactors}
\vspace*{10pt}
\centering
\begin{tabular}{lcc}
\hline
\multicolumn{1}{c}{Molecule} & Constant & Ref$^a$ \\
 & (ergs mol$^{-1}$ sec$^{-1}$) \\
\hline
OH & $f(\dot{R}_h)$ & 1 \\
NH & $f(\dot{R}_h)$ & 2 \\
CN & $f(\dot{R}_h)$ & 3 \\
C$_{3}$ & 12.42 & 4 \\
CH & 12.98 & 4 \\
C$_{2}$ ($\Delta v=1$) & 12.62 & 5 \\
C$_{2}$ ($\Delta v=0$) & 12.35 & 5 \\
NH$_2$ & 13.73 & 6 \\
\hline
\multicolumn{3}{l}{For OH, NH and CN} \\
\multicolumn{3}{c}{$log$ M = $f(\dot{R}_h)$ + $log$ L + 2 $log$ $R_h$} \\
\multicolumn{3}{l}{For C$_{3}$, CH, C$_{2}$, and NH$_{2}$:} \\
\multicolumn{3}{c}{$log$ M = const + $log$ L + 2 $log$ $R_h$} \\
\multicolumn{3}{p{3.5in}}{where M = column density, L = band intensity, R$_h$ =
heliocentric distance ({\sc {\sc au}}), $\dot{R}_h$ = heliocentric radial
velocity (km/sec), $f(\dot{R}_h)$ is the fluorescence efficiency
as a function of heliocentric distance.} \\ [5pt]
\hline
\multicolumn{3}{p{3.5in}}{$^a$References -- 1: $g-$factor from Schleicher and
A'Hearn 1982; 2: $g-$factor from Kim {\it et al.} 1989; 3: Tatum and Gillespie
(1977); 4: Cochran and Barker (1985); 5: Oliversen {\it et al.} (1985) and
$g-$factor from A'Hearn {\it et al.} (1985); 6: $g-$factor from Tegler and
Wyckoff (1989)} \\
\hline
\end{tabular}
\nocite{scah82,kiahco89,tagi77,coba85,olhobr85,ahbifemi85,tewy89}
\end{table*}
These assume the coma is optically thin.  This will always be the case in the
outer coma.  For a few of the most active comets, the collisional zone
might be big enough to affect the inner pixel.  However, our pixels are
generally much larger than the affected region.  One comet for which this
would be a problem if it had been observed near 1\,{\sc au} is comet
C/1995~O1 (Hale-Bopp).  However, as our observations were at 4\,{\sc au}, this
is not a problem for our data.

\begin{figure}[ht!]
\includegraphics[scale=0.32,angle=270]{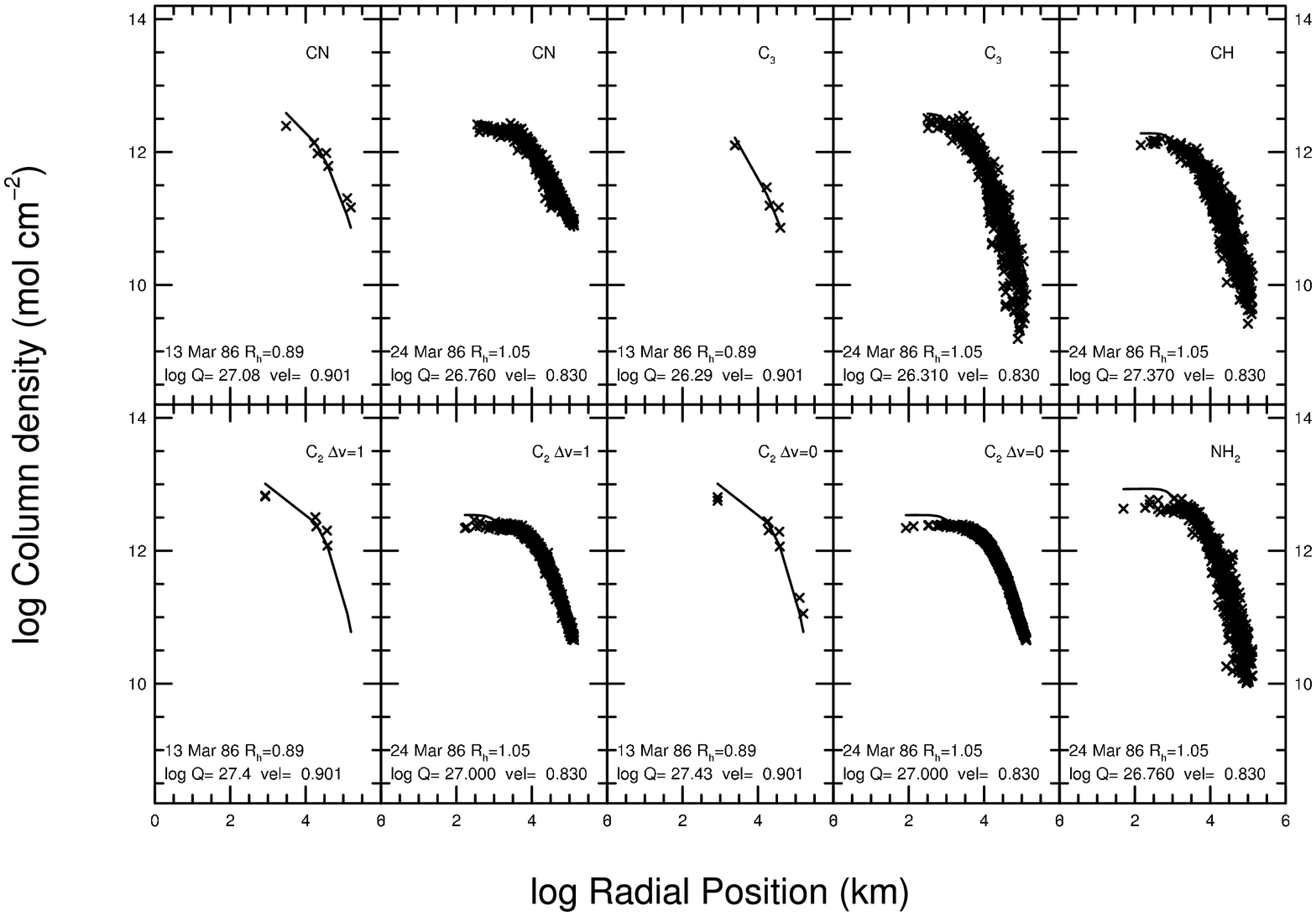}
\caption[fig3]{The column densities as a function of position in
the coma are shown for comet Halley.  In this figure, we show the
data sets from both the IDS (13 March 1986) and LCS (24 March 1986).  For
the IDS, we only observed 4 bands, while we observed 6 bands with the LCS.
Note the much larger number of data points obtained with the LCS, the result
of the long slit.  Superimposed on the data are the Haser model fits
to the data. The production rates, log Q, are in mol sec$^{-1}$; the
velocity, vel, is in km sec$^{-1}$; the heliocentric distance, R$_h$, in
in {\sc au}.}
\label{halley}
\end{figure}

Note that at
this point we generally had column densities for each observed molecule at
multiple positions within the coma.  For some comets, we might just have an
optocenter point, while for others we had many hundreds of data points,
covering distances to greater than 10$^5$\,km from the optocenter. 
Figure~\ref{halley} shows an example of the column densities as a function
of position for comet Halley obtained using both the LCS and IDS.
We needed a
way to be able to intercompare comets that had disparate numbers of observed
positions in the coma
and a variety of observed molecules.  We did this by modeling the comets
using the computationally simple Haser (1957) model.  \nocite{ha57}
The Haser model is not very physical as it assumes that the cometary comae are
spherically symmetric and that all gas only flows outwards from the nucleus.  As
we will show later, spherical symmetry is often violated.  We also expect
that dissociation will allow the daughter species to travel in any random
direction.  Despite these shortcomings, we still chose to use this model
because it is an easy way to intercompare comets and to compare with other
researchers.  We adopted the Haser model scale lengths listed in
Table~\ref{scalelengths} for this purpose. 
\begin{table*}[ht!]
\caption{Adopted Haser Model Scale Lengths} \label{scalelengths}
\vspace*{10pt}
\centering
\begin{tabular}{lccc}
\hline
& Parent & Daughter & \\
& Scale length & Scale length & \\
Molecule & (km) & (km) & Reference$^a$ \\
\hline
OH & 2.4$\times10^4$ & 1.6$\times10^5$ & 1 \\
NH & 5.0$\times10^4$ & 1.5$\times10^5$ & 2 \\
CN & 1.7$\times10^4$ & 3.0$\times10^5$ & 3 \\
C$_{3}$ & 3.1$\times10^3$ & 1.5$\times10^5$ & 3 \\
CH & 7.8$\times10^4$ & 4.8$\times10^3$ & 4 \\
C$_{2}$$^b$ & 2.5$\times10^4$ & 1.2$\times10^5$ & 3\\
NH$_{2}$ & 4.1$\times10^3$ & 6.2$\times10^4$ & 5  \\
\hline
\multicolumn{4}{p{3.5in}}{$^a$References -- 1: Cochran and Schleicher (1993);
2: Randall {\it et al.} (1992); 3: Cochran (1986); 4: Cochran and Cochran (1990); 5: Cochran {\it et al.} (1992)}\\
\multicolumn{4}{l}{$^b$ C$_{2}$ parent scales as R$_{h}^{2.5}$, all others as R$_{h}^{2}$} \\
\hline
\end{tabular}
\nocite{co86haser}
\nocite{coco90,cosc93,raetal92}
\end{table*}
These are not necessarily the
best fits to all of the data but offer a consistent set of parameters
without too many degrees of freedom. Our choice of scale lengths,
as compared with other authors, is discussed later.
In addition, we use a modified
Delsemme (1982) velocity scaling law of $v = 0.85 R_h^{0.5}$ for the
models.

Table~\ref{prodrates} is representative of the information that we
gathered for each comet. 
%\begin{landscape}
\begin{table*}
%\scriptsize
\tiny
\caption{Representative Production Rate Table (Broken into 3 parts)} \label{prodrates}
\vspace*{10pt}
\begin{tabular}{lr@{ }l@{ }lcccccccccc}    %14 columns
\hline
\multicolumn{1}{c}{Comet} & \multicolumn{3}{c}{Date}    & PA & Instr. & R$_h$ & $\dot{R}$ & $\Delta$ & Group & Phot.? & Log Q(OH) & Bin & Npts \\
 & & & & (deg) & & (AU) & (km/sec) & (AU) & & & (mol/sec) & & \\
\hline
23P/Brorsen-Metcalf & 16 & Jul & 1989 & 88 & LCS & 1.31 & -28.51 & 0.84 & HTC & n & 27.99 & 2 & 104 \\
 & 09 & Aug & 1989 & 175 & LCS & 0.90 & -29.68 & 0.63 &  &y&  &  &  \\
 & 10 & Aug & 1989 & 175 & LCS & 0.88 & -29.64 & 0.63 &  &n& 27.98 & 2 & 54 \\
 & 11 & Aug & 1989 & 190 & LCS & 0.86 & -29.58 & 0.63 &  &n& 28.60 & 2 & 225 \\
\hline \\
\end{tabular}
\begin{tabular}{cccccccccccc}   % 12columns
\hline
log Q(NH) & Bin & Npts & log Q(CN) & Bin & npts & log Q(C$_3$) & Bin & Npts & Log Q(CH) & Bin & Npts \\
(mol/sec) & & & (mol/sec) & & & (mol/sec) & & & (mol/sec) \\
\hline
25.37 & 1 & 148 &25.25 & 1 & 217 & 24.64 & 1 & 184 & 25.68 & 1 & 157 \\
26.11 & 1 & 467 &25.86 & 1 & 505 & 25.11 & 1 & 449 & 26.38 & 1 & 463 \\
26.13 & 1 & 175 & &  &  &  &  &  &  &  &  \\
26.17 & 1 & 503 & 25.83 & 1 & 505 & 25.19 & 1 & 389 & 26.34 & 1 & 354 \\
\hline \\
\end{tabular}
\begin{tabular}{ccccccccc}  % 9 columns
\hline
 Log Q(C$_2$) & Bin & Npts & Log Q(C$_2$) & Bin & Npts & Log Q(NH$_2$) & Bin & Npts     \\
(mol/sec) & & & (mol/sec) & & & (mol/sec) \\
 $\Delta v=1$ & & & $\Delta v=0$ \\
\hline
 25.47 & 1 & 213 & 25.46 & 1 & 217 &  &  &      \\
 26.10 & 1 & 490 & 26.09 & 1 & 496 & 25.35 & 1 & 277     \\
  &  &  &  &  &  &  &  &      \\
 26.10 & 1 & 498 & 26.11 & 1 & 504 &  &  &      \\
\hline
\end{tabular}
\end{table*}
%\end{landscape}
For each comet and each night of observation, we
list the observational circumstances, the instrument, the position angle
of the slit for the LCS and ES2, measured north through east
(left blank for the IDS since the apertures were fixed on an east/west line)
and the values of the measured production rates 
(denoted with a Q(X) where X is the species in question).
For the production rates, we list the log of the production rate for
each molecule along with the binning (for LCS and ES2) and the number
of data points that went into the fit.  Note that for
some comets and some nights we would have more than one spectral
image per night, resulting in many more data points than are covered
by the slit in a single spectral image.  These spectral images were
sometimes, but not always, centered on different portions of the coma.

The table here is for only
a single comet (23P/Brorsen-Metcalf) to give the reader an idea of
the nature of the available data.  
As it is, this representation of the table must be sliced into three parts to 
fit the table onto a page for printing.
Since there are over 650 lines of table with all of the comets, it
is impractical to print all of the values.  Instead, all of the data
are available on-line in the supplementary materials for this paper.

One of the columns in this table is labeled ``Group".  This column
lists the dynamical type of the comet: JFC = Jupiter Family comet (Tisserand
parameter $> 2$); HTC = Halley
Type Comet (Tisserand parameter $<2$ and P $<$ 200 years);
LPC = Long Period Comet (P $>$ 200 years).
The next column is labeled ``Phot?".  The values for this column are
either "y" or "n" to indicate whether or not it was photometric.  If it was
not photometric, then the production rates are not absolute.  However, since
all the wavelengths were observed simultaneously, the relative productions
of species are still accurate since most clouds are gray (Wing 1967).
\nocite{wi67}
For the LCS and ES2, we
would sometimes obtain multiple spectral images even if not photometric.
For purposes of the analysis, we would scale all the column densities
(as a group from a single spectral image) to the spectral image that was
the brightest.  For this scaling, all molecules were scaled by the same
amount. That we could use a single scaling factor to bring molecules
at disparate wavelengths into good alignment is another indicator that
the clouds are generally gray.
The resultant
production rates are still not absolute but, again, the relative
production of species is correct.

When we observed at more than one slit position angle with the LCS or ES2,
we would keep the different position angles separate for the analysis.
The position angle should not affect the production rates (as long as the
optocenter is in each image) so this serves as a check on the derived
production rates.  The exception to this policy was the Encke observations
of October 2003 as we used so many different position angles that it was
impractical to keep them separate.

Inspection of Table~\ref{prodrates} shows that some of the entries for
production rates are blank.  In the case of Brorsen-Metcalf, these blanks
are because the bandpass did not include some of the molecules on
some of the nights.  On 16 July 1989 and 11 August 1989, we observed with
our nominal setup that covered OH but not NH$_{2}$.  On, 9 August 1989,
we changed the grating angle to observe NH$_{2}$ but not OH.  On 10
August 1989, we employed a higher resolving power grating and only observed
OH and NH.

For other comets, there are blank entries even when we covered a molecule
in our setup.  This means that we did not detect the molecule, generally
because the data were too noisy.  We consider that we {\it should} be able
to measure CN, C$_{3}$, and C$_{2}$ in all comets (OH and NH are affected
by airmass and low detector response, CH is very weak and NH$_{2}$ is often
off the chip). 
We can put upper limits on the column densities of these three
molecules by looking at how
the data in the bandpass of the emission band deviates from the continuum.
To measure an upper limit, we first fit a continuum to the first and
last three points in the bandpass.  Next, we computed a standard deviation,
$\sigma = sqrt [ \Sigma(observed - continuum)^2 / (N - 1)] $, where $N$
is the number of data points in the bandpass.  The upper limit is then 
defined as $U.L. = 0.75 \sigma B$, where $B$ = the bandpass in \AA.  This is
a 3$\sigma$ upper limit.

\begin{figure}
\includegraphics[scale=0.33,angle=270]{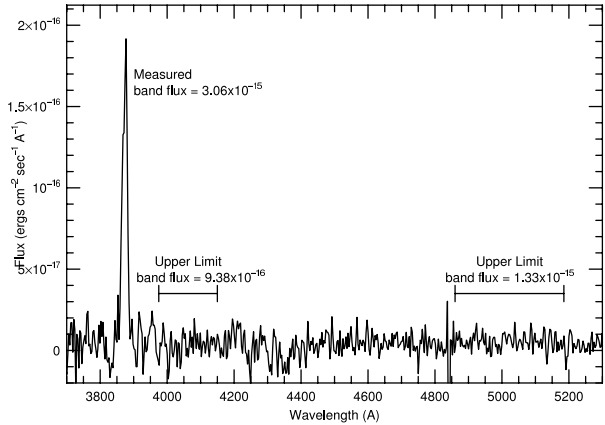}
\caption[fig4]{A spectrum of comet 192P/Shoemaker-Levy is shown.
CN is obviously visible but there is no sign of the C$_{3}$ or C$_{2}$.
By measuring the noise in the bandpass where we would expect to detect
these features, we can derive upper limits for the amount of gas that
could be attributable to these species.  These upper limits are $3\sigma$
limits as discussed in the text. 
}
\label{upper}
\end{figure}

Figure~\ref{upper} shows an example of a comet for which we detected
the CN band but there is no obvious C$_{3}$ or C$_{2}$. 
This spectrum, of
comet 192P/Shoemaker-Levy at R$_h$ of 1.84\,{\sc au}, has had
the telluric and solar spectra removed.  The bandpasses for the C$_{3}$ and
C$_{2}$ $\Delta v=0$ bands are marked on the figure.  Included are the
measured integrated band intensity for CN along with the $3\sigma$ upper limit
integrated fluxes for C$_{3}$ and C$_{2}$.  In this example, it should be
obvious that there cannot be much C$_{2}$ or C$_{3}$ hidden in this
spectrum.

In general, the upper limits are not very constraining.  If we observed
all three of CN, C$_{3}$, and C$_{2}$ on at least one of the nights we
observed a particular comet, we did not tabulate the upper limits 
for other nights because we believed that
the actual detections were more meaningful (the upper limit measurements are
much more labor intensive than the usual column density measurements
because we never developed the same degree of automation in their computation
that we have for column densities).
However, we have included upper limits for the cases when CN was detected
but C$_{2}$ and C$_{3}$ were not both detected.  We have only one case,
comet C/1984\,K1 (Shoemaker), when we detected CN and C$_{3}$ but not
C$_{2}$ and one case, C/1992 W1 (Oshita), where we observed CN and C$_{2}$ but
not C$_{3}$.

Of our sample of 130 comets, there are 20 comets for which we observed 
\underline{no} molecules, not even CN.  These are tabulated in
Table~\ref{no_mols}, where we list the date, observing circumstances,
group and the upper limits for the CN production rates. 
\begin{table*}
\caption{Comets With No Molecules Detected} \label{no_mols}
\vspace*{5pt}
\centering
\setlength{\tabcolsep}{1.5pt}
\begin{tabular}{lr@{\,}l@{\,}lcccccc}
\hline
\multicolumn{1}{c}{Comet} & \multicolumn{3}{c}{Date} & Instr. & R$_h$ &
$\dot{R}$ & $\Delta$ & Group & Log Q(CN) \\
 & & & & & (AU) & (km/sec) & (AU) & & Upper Limit \\
 & & & & &      &          &      & & (mol/sec) \\
\hline
47P/Ashbrook-Jackson & 16 & Oct & 1985 & IDS & 2.41 & -3.52 & 2.16 & JFC & $<$24.04 \\
 & 16 & Nov & 1985 & IDS & 2.36 & -2.54 & 2.47 &  & $<$24.07 \\
 &  &  &  &  &  &  &  &  &  \\ [-5pt]
C/1980 E1 (Bowell) & 12 & Dec & 1980 & IDS & 5.38 & -11.85 & 5.56 & LPC & $<$25.64 \\
 & 15 & Dec & 1980 & IDS & 5.36 & -11.83 & 5.49 &  & $<$25.29 \\
 & 03 & Jan & 1981 & IDS & 5.23 & -11.65 & 5.07 &  & $<$25.19 \\
 & 07 & Feb & 1981 & IDS & 5.00 & -11.29 & 4.33 &  & $<$25.71 \\
 & 11 & Feb & 1981 & IDS & 4.98 & -11.25 & 4.26 &  & $<$25.56 \\
 & 07 & Jun & 1981 & IDS & 4.27 & -9.72 & 3.88 &  & $<$24.91 \\
 & 24 & Jan & 1982 & IDS & 3.39 & -2.24 & 3.92 &  & $<$25.67 \\
 & 25 & Jan & 1982 & IDS & 3.39 & -2.19 & 3.90 &  & $<$25.41 \\
 & 20 & Apr & 1982 & IDS & 3.38 & 1.84 & 2.82 &  & $<$24.90 \\
 & 18 & May & 1982 & IDS & 3.42 & 3.12 & 2.57 &  & $<$25.25 \\
 & 26 & Jun & 1982 & IDS & 3.51 & 4.77 & 2.50 &  & $<$25.40 \\
 & 16 & Jul & 1982 & IDS & 3.57 & 5.55 & 2.61 &  & $<$25.21 \\
 & 17 & Jul & 1982 & IDS & 3.58 & 5.59 & 2.62 &  & $<$25.17 \\
 & 18 & Jul & 1982 & IDS & 3.58 & 5.63 & 2.63 &  & $<$25.11 \\
 & 17 & Aug & 1982 & IDS & 3.69 & 6.68 & 2.99 &  & $<$25.14 \\
 & 15 & Sep & 1982 & IDS & 3.81 & 7.58 & 3.47 &  & $<$25.15 \\
 & 13 & Oct & 1982 & IDS & 3.94 & 8.33 & 3.99 &  & $<$25.06 \\
 & 10 & Jun & 1983 & IDS & 5.36 & 11.50 & 5.01 &  & $<$25.33 \\
 & 13 & Jun & 1983 & IDS & 5.38 & 11.52 & 4.99 &  & $<$25.08 \\
 & 07 & Jul & 1983 & IDS & 5.55 & 11.63 & 4.83 &  & $<$25.30 \\
 & 05 & Aug & 1983 & IDS & 5.74 & 11.73 & 4.78 &  & $<$25.07 \\
 & 10 & Sep & 1983 & IDS & 5.99 & 11.83 & 5.02 &  & $<$25.10 \\
 & 05 & Oct & 1983 & IDS & 6.16 & 11.88 & 5.41 &  & $<$25.11 \\
 &  &  &  &  &  &  &  &  &  \\ [-5pt]
C/1981 H1 (Bus) & 06 & Jun & 1981 & IDS & 2.53 & -4.46 & 1.91 & LPC & $<$23.73 \\
 & 07 & Jun & 1981 & IDS & 2.52 & -4.39 & 1.92 &  & $<$23.63 \\
 &  &  &  &  &  &  &  &  &  \\ [-5pt]
\hline
\end{tabular}
\end{table*}
\begin{table*}
\begin{center} Table~\ref{no_mols}: Comets With No Molecules Detected (cont.) \end{center}
\vspace*{5pt}
\centering
\setlength{\tabcolsep}{1.5pt}
\begin{tabular}{lr@{\,}l@{\,}lcccccc}
\hline
\multicolumn{1}{c}{Comet} & \multicolumn{3}{c}{Date} & Instr. & R$_h$ &
$\dot{R}$ & $\Delta$ & Group & Log Q(CN) \\
 & & & & & (AU) & (km/sec) & (AU) & & Upper Limit \\
 & & & & &      &          &      & & (mol/sec) \\
\hline
108P/Ciffreo & 12 & Dec & 1985 & IDS & 1.75 & 3.94 & 0.79 & JFC & $<$23.32 \\
 &  &  &  &  &  &  &  &  &  \\ [-5pt]
57P/duToit-Neujmin-Delporte & 01 & Jul & 1989 & LCS & 1.98 & -7.46 & 1.19 & JFC & $<$23.39 \\
 &  &  &  &  &  &  &  &  &  \\ [-5pt]
C/1983 N1 (IRAS) & 02 & Apr & 1984 & IDS & 4.24 & 13.40 & 3.27 & LPC & $<$24.49 \\
 &  &  &  &  &  &  &  &  &  \\ [-5pt]
48P/Johnson & 08 & Jul & 1983 & IDS & 2.51 & -4.45 & 1.50 & JFC & $<$23.61 \\
 & 06 & Aug & 1983 & IDS & 2.44 & -3.78 & 1.58 &  & $<$23.80 \\
 & 07 & Aug & 1983 & IDS & 2.44 & -3.76 & 1.59 &  & $<$23.21 \\
 &  &  &  &  &  &  &  &  &  \\ [-5pt]
C/1999 S4 (LINEAR) & 25 & Jan  & 2000 & LCS & 2.97 & -21.07 & 2.80 & LPC & $<$24.97 \\
 & 26 & Jan  & 2000 & LCS & 2.96 & -21.10 & 2.81 &  & $<$24.45 \\
 &  &  &  &  &  &  &  &  &  \\ [-5pt]
77P/Longmore & 04 & Jan & 1981 & IDS & 2.98 & -5.53 & 2.31 & JFC & $<$23.17 \\
 & 17 & Apr & 1988 & IDS & 2.65 & -4.39 & 2.09 &  & $<$23.23 \\
 &  &  &  &  &  &  &  &  &  \\ [-5pt]
96P/Machholz 1 & 13 & Jun & 2007 & LCS & 1.62 & 26.78 & 0.67 & JFC & $<$23.68 \\
 & 14 & Jun & 2007 & LCS & 1.64 & 26.62 & 0.68 &  & $<$23.43 \\
 &  &  &  &  &  &  &  &  &  \\ [-5pt]
7P/Pons-Winnecke & 06 & Apr & 1989 & IDS & 1.98 & -13.12 & 1.19 & JFC & $<$23.55 \\
 & 09 & May & 1989 & IDS & 1.73 & -12.56 & 1.17 &  & $<$23.76 \\
 &  &  &  &  &  &  &  &  &  \\ [-5pt]
30P/Reinmuth 1 & 18 & Nov & 1987 & IDS & 2.38 & -8.35 & 1.43 & JFC & $<$24.26 \\
 & 19 & Nov & 1987 & IDS & 2.38 & -8.34 & 1.42 &  & $<$23.81 \\
 & 24 & Dec & 1987 & IDS & 2.22 & -7.57 & 1.31 &  & $<$24.42 \\
 &  &  &  &  &  &  &  &  &  \\ [-5pt]
94P/Russell 4 & 04 & Jun & 1984 & IDS & 2.38 & 5.24 & 1.70 & JFC & $<$24.25 \\
 &  &  &  &  &  &  &  &  &  \\ [-5pt]
C/1983 R1 (Shoemaker) & 05 & Oct & 1983 & IDS & 3.38 & -2.25 & 2.45 & LPC & $<$24.39 \\
 & 06 & Dec & 1983 & IDS & 3.35 & 0.56 & 3.46 &  & $<$24.48 \\
 &  &  &  &  &  &  &  &  &  \\ [-5pt]
\hline
\end{tabular}
\end{table*}
\begin{table*}
\begin{center} Table~\ref{no_mols}: Comets With No Molecules Detected (cont.) \end{center}
\vspace*{5pt}
\centering
\setlength{\tabcolsep}{1.5pt}
\begin{tabular}{lr@{\,}l@{\,}lcccccc}
\hline
\multicolumn{1}{c}{Comet} & \multicolumn{3}{c}{Date} & Instr. & R$_h$ &
$\dot{R}$ & $\Delta$ & Group & Log Q(CN) \\
 & & & & & (AU) & (km/sec) & (AU) & & Upper Limit \\
 & & & & &      &          &      & & (mol/sec) \\
\hline
C/1989 A5 (Shoemaker) & 07 & Feb & 1989 & IDS & 2.65 & -1.39 & 1.70 & LPC & $<$25.17 \\
 & 08 & Feb & 1989 & IDS & 2.65 & -1.32 & 1.70 &  & $<$24.70 \\
 & 04 & Apr & 1989 & IDS & 2.67 & 2.66 & 2.43 &  & $<$24.04 \\
 & 08 & May & 1989 & IDS & 2.74 & 4.93 & 3.00 &  & $<$24.68 \\
 &  &  &  &  &  &  &  &  &  \\ [-5pt]
128P/Shoemaker--Holt 1 & 18 & Nov & 1987 & IDS & 3.22 & -2.96 & 2.40 & JFC & $<$24.58 \\
 & 11 & Dec & 1988 & IDS & 3.25 & 3.15 & 2.42 &  & $<$24.43 \\
 &  &  &  &  &  &  &  &  &  \\ [-5pt]
121P/Shoemaker-Holt 2 & 05 & Apr & 1989 & IDS & 3.01 & 4.56 & 2.27 & JFC & $<$24.15 \\
 &  &  &  &  &  &  &  &  &  \\ [-5pt]
56P/Slaughter-Burnham & 25 & Aug & 1981 & IDS & 2.63 & -3.25 & 1.87 & JFC & $<$23.30 \\
 & 26 & Aug & 1981 & IDS & 2.63 & -3.22 & 1.86 &  & $<$23.36 \\
 & 28 & Aug & 1981 & IDS & 2.62 & -3.15 & 1.84 &  & $<$23.56 \\
 & 27 & Sep & 1981 & IDS & 2.58 & -2.07 & 1.61 &  & $<$23.50 \\
 & 28 & Sep & 1981 & IDS & 2.58 & -2.04 & 1.60 &  & $<$23.47 \\
 & 24 & Oct & 1981 & IDS & 2.55 & -1.03 & 1.58 &  & $<$23.81 \\
 & 23 & Jan & 1982 & IDS & 2.59 & 2.54 & 2.51 &  & $<$23.81 \\
 & 25 & Jan & 1982 & IDS & 2.60 & 2.58 & 2.52 &  & $<$23.45 \\
 &  &  &  &  &  &  &  &  &                         \\ [-5pt]
74P/Smirnova-Chernykh & 26 & Oct & 1981 & IDS & 4.34 & -1.98 & 3.44 & JFC & $<$24.27                        \\
 & 13 & Oct & 1982 & IDS & 3.91 & -2.07 & 3.79 &  & $<$24.32                        \\
 & 14 & Oct & 1982 & IDS & 3.91 & -2.07 & 3.78 &  & $<$23.98                        \\
 & 16 & Dec & 1982 & IDS & 3.84 & -1.95 & 2.92 &  & $<$24.03                        \\
 & 06 & Dec & 1983 & IDS & 3.57 & -0.45 & 3.58 &  & $<$24.21                        \\
 & 08 & Dec & 1983 & IDS & 3.57 & -0.44 & 3.55 &  & $<$24.14                        \\
 & 25 & Feb & 1984 & IDS & 3.56 & 0.02 & 2.61 &  & $<$23.93                        \\
 & 31 & Mar & 1984 & IDS & 3.56 & 0.23 & 2.62 &  & $<$24.46                        \\
 &  &  &  &  &  &  &  &  &                         \\ [-5pt]
125P/Spacewatch & 15 & Jun & 1996 & LCS & 1.57 & -3.12 & 1.08 & JFC & $<$23.65                        \\
\hline
\end{tabular}
\end{table*}
We do not include
any other upper limits since ratios of upper limits are not at all
meaningful.  Inspection of Table~\ref{no_mols} shows that many of
these comets were observed at large heliocentric distance ($>$ 2.5\,{\sc au}),
where they would not be expected to produce much emission.  However, there
are a few examples (108P/Ciffreo, 96P/Machholz~1, 7P/Pons-Winnecke, and
125P/Spacewatch) with observations at $\leq 1.75$\,{\sc au}, where we
would have expected to detect at least CN.  These comets must be
very low producers, indeed.

Before we can compare comets, it is important to understand the uncertainties
in our measurements.  As with all data, these include both systematic
and random errors.  One source of the systematic errors is our 
knowledge of the $g$-factors. We differ from A'Hearn {\it et al.} (1995) for 
the C$_{3}$ and CN $g$-factors.  
The different $g$-factors
will make it hard to intercompare with other observers but should not
affect intercomparisons of comets within a single data set.

For C$_{3}$, no good oscillator strengths have been measured in the laboratory
since C$_{3}$ is not a stable molecule.  Jorgensen {\it et al.}. (1989) 
reviewed the laboratory measurements and found a range in measured
oscillator strengths of a factor of 6.  
\nocite{joalsi89, clla82,roetal02,adetal03}
We long ago adopted a value of $f = 0.016$ (Clegg and Lambert 1982) while
A'Hearn {\it et al.} use 0.001.  Roueff {\it et al.}. (2002) adopted a value of $f=0.0146$ whereas \'{A}d\'{a}mkovics {\it et al.} (2003) adopted
$f=0.016$, in agreement with our value.  Our value is consistent
with the current literature.  However, the oscillator strength is, in essence,
a scaling factor for the column densities, so our C$_{3}$ column densities
would be different by a constant factor from those computed with a
different oscillator strength.

CN is a bit more complicated than C$_{3}$ because one needs to account for
the Swings effect.  Thus, the fluorescence efficiency is a function of the
heliocentric radial velocity of the comet.  We have adopted the older
Tatum and Gillespie (1977) calculation of the fluorescence efficiency while
A'Hearn {\it et al.} use a formulation that factors in the heliocentric
distance of the comet.  Recent calculations by Schleicher (2010)
\nocite{sc10} 
are more complete, covering more heliocentric distances than previous
calculations.  Schleicher concludes that the values at a particular
heliocentric distance vary by about a factor of two with velocity but that
the trend of the fluorescence efficiency with radial velocity
at different heliocentric distances can be radically different.
Schleicher's calculations were published after we had completed the reduction
and analysis of much of our data.  We did not switch fluorescence efficiencies
because of the large investment in manpower at that point.  The effect
of using the older calculation is not a simple constant offset, as 
with the C$_{3}$ oscillator strength.  It will be less than the full
range of variation for the fluorescence efficiency, or a factor of 2 (0.3
in the log), and will be responsible for increasing the error bars when
values are averaged later in this paper.

Another source of systematic errors is our knowledge of the scale lengths
for the model.  We do not take into account any change in the strength
of the solar UV spectrum nor do we take into account any multi-step 
dissociation processes.  Inspection of the model fits in Figure~\ref{halley}
show examples of good fits (CN and C$_{3}$) and poor fits (C$_{2}$ and
NH$_{2}$).  The C$_{2}$ is almost always a bad fit in the very inner coma 
because
C$_{2}$ is probably a granddaughter product of its source and the Haser model
does not account for the two-step production.  In addition, some of these
scale lengths work for some, but not all, comets.  This problem was
noted by LS11 who chose to fit the scale lengths
for each comet.  We do not do this because then there are too many free
parameters in the intercomparison of comets.  Indeed, if the UV insolation
does not change, the scale lengths are, to first approximation, determined
entirely by the solar flux and the distance of the object from the Sun.
While collisions might affect the inner coma, the vast majority of the
coma is unaffected.  Our fits rarely end up being dependent
on the inner coma data points, anyway, since we have so many data points
at larger cometocentric distances. We described the process of defining the
scale lengths in Cochran (1986) \nocite{co86haser} and use these or
updated scale lengths {\it uniformly} for all of our observations.
These scale lengths were defined using observations from five comets
observed from 0.74 to 1.81\,{\sc au}.
We include an expanded discussion of the scale lengths later when we
intercompare different data sets.

\begin{figure}[ht!]
\includegraphics[scale=0.45]{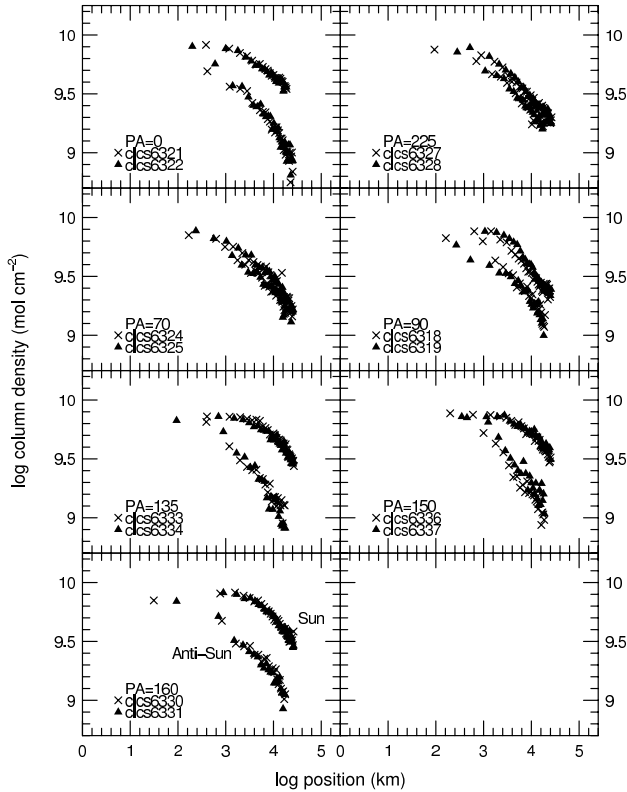}
\caption[fig1]{The CN column densities for 2P/Encke observed
on 24 October 2003 are shown.  Each panel represents a different slit
position angle (and that position angle $\pm 180^\circ$ on the other side
of the optocenter). There are two spectral images at each position angle.
Clearly, comet Encke's coma is not symmetric. 
}
\label{encke}
\end{figure}

Probably the most important source of systematic error is the Haser model
itself.  This model assumes that the cometary coma is spherically symmetric
and that gas can only flow outwards.  With our long slit observations, we
observe that spherical symmetry is rarely a good approximation. The
effect of using the Haser model is, in affect, to azimuthally average the 
coma observations and assume this is a reasonable proxy for the
production.  Figure~\ref{encke} is an (albeit extreme) example of the 
types of asymmetry we observe. 
In these observations of comet 2P/Encke,
obtained on 24 October 2003, we rotated the slit to 7 different position
angles.  Since
the comet was somewhere near the center of the detector, spatially, 
we actually obtained data on two sides of the
optocenter (the listed position angle and the position
angle $\pm$ 180$^\circ$) for each listed position angle.  
Additionally, at each position angle we obtained two spectral images (that are
shown with different symbols in each panel).  Inspection of this figure
shows that Encke's coma is far from symmetric! The asymmetries observed are not
due to clouds or other errors.  This is demonstrated by the
fact that both spectral images at a single position angle are in
excellent agreement (no scaling was done).
At some
position angles (e.g. 70$^\circ$, when the slit was positioned perpendicular
to the Sun/tail line), the coma looks reasonably symmetric,
while at others (e.g. 160$^\circ$, along the Sun/tail line),
there is a factor of 10 difference
in column density by the time we are a few thousand km from the optocenter.
A simple model like the Haser model cannot do these data justice.  A
more complicated model, such as proposed by Ihalawela {\it et al.} (2011),
\nocite{ihpidoco11}
allows for deviations from spherical symmetry.  However, it should
be noted that if we independently fit the different position angles
in these Encke data with the Haser model 
(using both sides of the optocenter together in a single fit),
we get the same CN production rate for each position angle
(even though the fits look bad).
Asymmetries are observed in many comets.  Combi and Fink (1993) showed that
Haser model scale lengths could be derived to accommodate time 
variable asymmetries in comet Halley.
\nocite{cofi93}

Another source of potential systematic error is the flux calibration.  The
fluxes of the standard stars are probably only good to 5--10\%.  In addition,
though we used a large aperture for observations of the standard stars, 
there could be a loss of light from the aperture on some nights.  This
would cause a decrease in the amount of light we observed for the star and
our sensitivity function would then cause a slightly higher flux for
the cometary spectra.  This would be an extremely rare occurrence, especially
for the LCS, since we used a 10\,arcsec wide slit.

Clouds would also cause a diminution of the flux.
They can affect the comet and standard stars by differing amounts.
Thus, it could cause the cometary
flux to appear brighter or dimmer.  As noted above, clouds are
gray and we observe all wavelengths at the same time, so we can
intercompare the different bands in a ratio sense.  In addition, with the
long slit, we observe many different positions at the same time, so the
column densities as a function of position maintain an accurate relationship. 
It is only when we were obtaining more than one spectral image that we had
to scale the data.  As noted above, we would use the same scale factor at
all wavelengths.

An additional source of error was continuum removal.  This is required in
order to study the gas. Sometimes the noise in the continuum meant that
the process was not as accurate as we would desire.  We mitigate this 
uncertainty in large part by fitting a continuum under the band prior
to computing the integrated band intensity.  This corrects for any
continuum removal errors.

Finally, an important stochastic error is the photon statistics.  For
some of our observations, such as the Encke data shown in Figure~\ref{encke},
it is clear that the signal/noise was very high.  For others, the 
signal/noise is not at all good.  This is especially true for some of
the IDS data where we chose to obtain spectra of another object instead
of building signal on the object we were observing.  As is normal for stochastic
errors in data, we can lower the impact of this noise by collecting more
data, i.e. by observing more positions in the coma.  For our long
slit CCD data, this is a natural consequence of the long slit.  

We can judge the affect of the noise on our answer by using a
Monte Carlo model to simulate changing the noise.  Photon statistical
errors affect each point individually.  Thus, for a given signal,
the measured value could vary based on the signal/noise.  To simulate
the effect, we took a data set and we changed the value of each data point
by selecting randomly from a set of normal Gaussian deviates. Once each
data point was modified, we fit the Haser model and recorded the value for
the production rate.  We then did this same process repeatedly.  Finally,
we figured the mean and standard deviations of the derived column densities
(in linear space) with the model that was fit to the real data. 
We tested
the errors by running the simulation with 1000 sets of modified data.
\begin{figure}
\includegraphics[scale=0.33,angle=270]{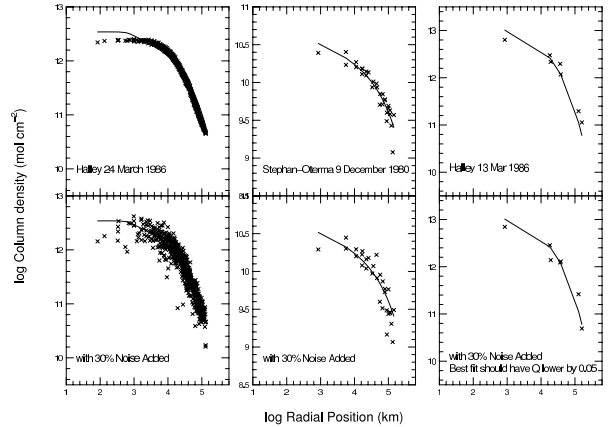}
\caption[fig6]{The effects on the column densities due to
photon noise are illustrated in these panels.  See the text for a
discussion of the model.  The top panels are three different C$_{2}$ data sets
with 433 individual column densities (Halley 24 March), 30 column
densities (Stephan-Oterma 9 December) and only 7 column densities (Halley
13 March).  In the top panels, the original data are shown with the
best fit Haser model.  In the bottom panels, the data are shown after we draw
new values from a normalized Gaussian distribution for each point based on
30\% potential error.  The model fit is the same one as
the real data and looks good for all three data sets.  However, while
the same model is correct for the Halley 24 March and Stephan-Oterma data,
the Halley 13 March data actually required lowering the log
C$_{2}$ column density by 0.05.  This change in best fit is based on
1000 simulations of noise.
}
\label{noise}
\end{figure}
The range of values that can be selected are generally set by the
signal/noise of the data.  Figure~\ref{noise} shows examples of the
process in three different C$_{2}$ data sets - one with 433 data points (Halley
from 24 March 1986), one with 30 data points (Stephan-Oterma from
9 December 1980) and one with 7 data points (Halley from 13 March 1986).
In all three cases, we added 30\% noise, though it is apparent from the
scatter in the data that 30\% noise is an over estimate, at least for
the 24 March Halley  (actual noise about 2.5\%) and for Stephan-Oterma 
(noise of about 7\%).  As shown in the figure,
even with the added noise, the Haser model gives the same values
for the production rates for those two data sets. 
For the 13 March Halley data, 30\%
error would result in a change of the production rate of only 0.05 in the
log.  Thus, to the extent that the model and the scale lengths are 
appropriate, our column densities are very robust when we have
more than 25 data points.  Fewer data points means that the uncertainty
in the production rates might be 5 -- 20\% (stochastically).  This is
why the number of data points is included in the table that is
available on line and shown as an example in Table~\ref{prodrates}.

\begin{figure}[ht!]
\includegraphics[scale=0.45]{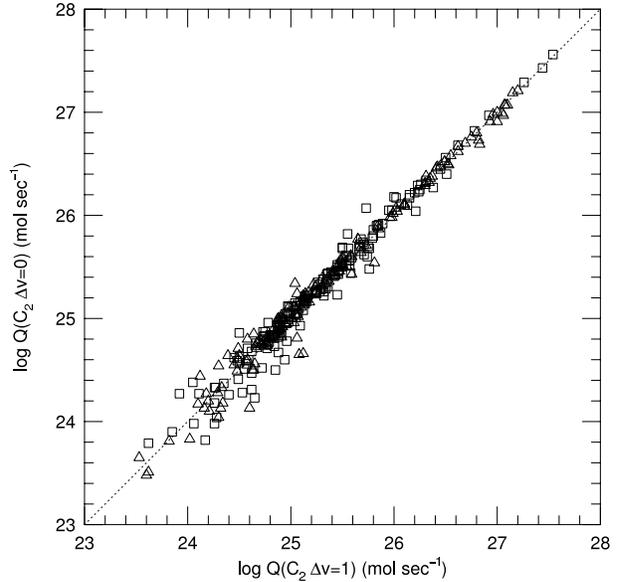}
\caption[fig7]{
The measured values of the production
rates for the C$_{2}$ $\Delta v=1$ and C$_{2}$ $\Delta v=0$ bands are shown
plotted against one another.  We would expect to derive the same production
rates from either band.  This is what we observe.  The IDS data
are denoted with squares while the LCS data are denoted with triangles.
The uncertainties in the IDS production rates are generally higher
than for the LCS production rates since we observe so many
more positions with the LCS.  The disagreement of the
production rates increases at the smallest production rates, indicative
of larger uncertainties for weaker comets. }
\label{C2}
\end{figure}

In addition to looking at the number of data points and how much
scatter appears in plots of column density versus position, one can check
consistency using C$_{2}$ observations. 
We measure both the C$_{2}$ $\Delta v=0$ and
$\Delta v=1$ bands in our spectra.   The $\Delta v=0$ band
has a larger integrated band intensity than the $\Delta v=1$ band.
This is accounted for by the different constants used for conversion
of the intensities to the column densities.
Therefore,
the two bands should give {\it independent} measurements of the C$_{2}$
column densities.  Figure~\ref{C2} shows our derived column densities
for all comets on all nights on which we measured both bands.  There
should be a 1:1 correlation. 
Inspection of Figure~\ref{C2} shows that
in general this is true, though there are some outliers. 
We use different symbols for the IDS and LCS data.
Most of the outliers are the IDS data, as would be expected from the 
much smaller numbers of points obtained for the IDS, resulting in
larger error bars on the derived production rates.  
In particular,
we see the greatest deviations from the correlation at the smallest
values for the column densities.  This is not surprising as the weakest
features are the ones with the largest uncertainties.  When log Q for
either band is $>$25, the agreement is generally quite good.  
A few outliers with log~Q~$> 25$ exist, and were checked, but there was nothing to pick one
value over the other.

\section{Trends in the data}

\begin{figure}
\includegraphics[scale=0.33,angle=270]{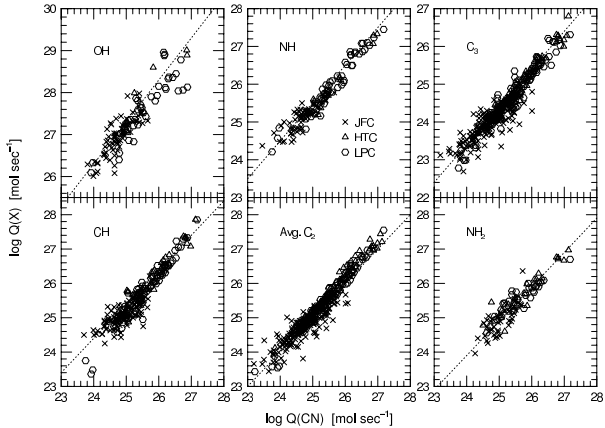}
\caption[fig8]{This figure illustrates the observed production
rate trends.  In each panel, we show the log of the production rate of
a particular molecule plotted against the log of the CN production rate.
A slope equal 1 line is included on the plot to guide the eye (these lines
are not fits).  We see that the production rates of all of the molecules
correlate well with the CN production rates.  Except that JFCs are generally
lower gas producers, we see no obvious trends with dynamical type.
}
\label{molvsmol}
\end{figure}

Figure~\ref{molvsmol} shows the observed column densities for all
comets on each night they were observed, with
OH, NH, C$_{3}$, CH, C$_{2}$ and NH$_{2}$ being plotted against CN.  
While OH, a daughter of the most abundant ice H$_{2}$O,
would be the most desirable molecule to use to compare one comet against
another, we chose to plot abundances versus CN because CN is the most
commonly observed feature in our spectra.  Inspection of this plot shows
that the production rates of each species, including OH, are well correlated
with the production rate of CN.  Thus, comparing to CN is a reasonable approach.
In this figure, we denote the different dynamical types of comets by
different symbols.  We see no obvious trend with dynamical type except
that JFCs tend to produce less of all species.  HTCs and LPCs seem to
produce comparable amounts of gas.

Eighteen of the comets were observed on only one night each. 
An additional 20 comets were only
observed on two nights each.  The remaining 72 comets were each observed on
at least
three different nights or position angles, with comet 9P/Tempel 1 having
45 observations.  Clearly then, some comets are dominating the trends
in Figure~\ref{molvsmol}.   We can mitigate this effect by combining
all the data for an individual comet.

\begin{figure}[ht!]
\includegraphics[scale=0.33,angle=270]{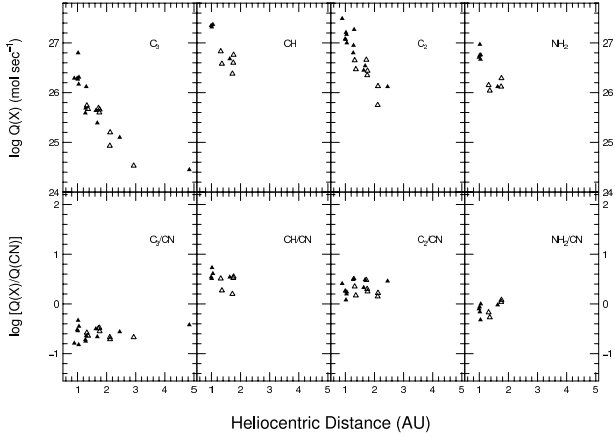}
\caption[fig9]{This figure shows how the gas production
rates change as a function of heliocentric distance (upper panels) for
comet Halley.  We
see each species showing increased production at smaller heliocentric
distances.  Closed symbols denote photometric nights while open symbols
mean it was not photometric.  In the lower panels we plot the ratio of
the production rate of each species to that of CN, with the y axis
having the same overall range (3.8 dex) as in the upper panels.  Notice that the
production rate ratios are constant with heliocentric distance. 
}
\label{hal_trends}
\end{figure}

In order to combine the data, we first took ratios of the production
rates of each species with respect to CN, for each date
when both molecules were detected.  This had the effect of
removing any weather-related differences between comets, as clouds
would affect all species equally.  In addition, unless a comet
is inhomogeneous, we expect that all the species should be produced
in the same ratio at different heliocentric distances.  Figure~\ref{hal_trends}
shows a comparison of the change in production rates as a function
of heliocentric distance versus the change in production rate ratios
as a function of heliocentric distance for comet 1P/Halley. 
The production rates show
a dramatic increase when the comet is at smaller
heliocentric distances.
In contrast, the production rate ratios are constant
with heliocentric distance. 
Fink (1994) also found a constant mixing ratio with heliocentric
distance for comet Halley. \nocite{fi94}

\begin{figure}
\includegraphics[scale=0.33,angle=270]{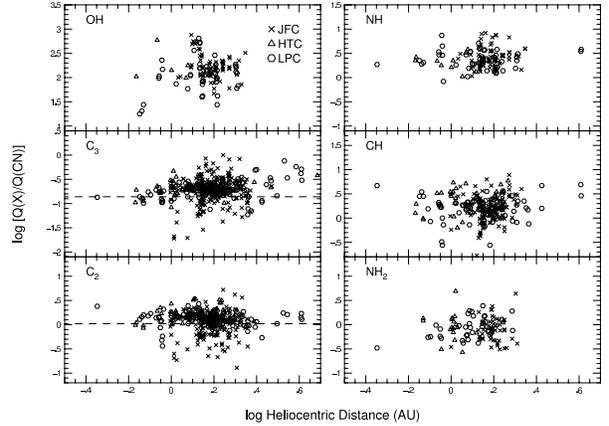}
\caption[fig10]{The production rate ratios are plotted as a
function of log heliocentric distance for all instances where we observed,
ratioed to CN for the same night.  Each point represents one comet
on one particular night.
The different dynamical types are denoted with different symbols with the
symbols defined in the OH panel.  There
are no obvious trends with heliocentric distance.  The dashed lines for
C$_{2}$ and C$_{3}$ represent the dividing line between carbon-chain typical
and depleted comets (see text). 
}
\label{helio}
\end{figure}

Figure~\ref{helio} shows the observed production rate ratios as
a function of log heliocentric distance (using the log for the heliocentric
distance just spreads out the data along the x axis since the vast majority
of data were obtained between 1 and 2.5\,{\sc au}). 
With the possible exception of C$_{3}$, there are not any
obvious trends with heliocentric distance, as would be expected if
the species are outgassing together.  We also see no obvious trends
with dynamical type. 
The dashed lines in the C$_{2}$ and C$_{3}$ 
panels denote our delimiters for depleted comets, discussed below.

The C$_{3}$ ratios for log heliocentric distance $> 0.5$ appear to be
systematically higher than the rest.  There are 8 data points at these large
distances but the data represent only 5 comets (there are three Hale-Bopp
points).  One of those is comet 1P/Halley. Inspection of Figure~\ref{hal_trends}
suggests this point is consistent with other Halley observations at 
smaller heliocentric distances.  Comet C/1983 O1 (Cernis) has the highest
C$_{3}$ ratio at this distance. This is the only C$_{3}$ value measured
for Cernis so
we do not know if the comet is different or the value is high because
of the large heliocentric distance.
Comet C/1993 A1 (Mueller) has the next highest value but is also on
the high side at 2\,{\sc au} ($log$ distance = 0.31). Comet C/1996 P1 (Wilson)
has a high value of C$_{3}$ at 3.5\,{\sc au} but normal-to-low
values at 1.34 and 3.13\,{\sc au}, suggesting the high value at 3.5\,{\sc au}
may be in error.  Three of the points belong to C/1995 O1 (Hale-Bopp).
We only observed this comet at large heliocentric distances as part of
this project so we do not know if the comet is different or there is
a heliocentric distance effect.

The constancy of production rate ratios with heliocentric distance
means that we can compute 
average values for the ratios that are representative of each comet
at all observed heliocentric distances.  The averages make
the data for a comet with 3 observations as meaningful as a comet
with 40 observations.
In addition, since all species were observed simultaneously, the effects
of clouds are canceled.
Similarly, we have computed an average CN production for each
comet by computing the production the comet would have at 1\,{\sc au}
assuming a 1/R$_h^2$ dependence of the production.

\begin{figure}
\includegraphics[scale=0.33,angle=270]{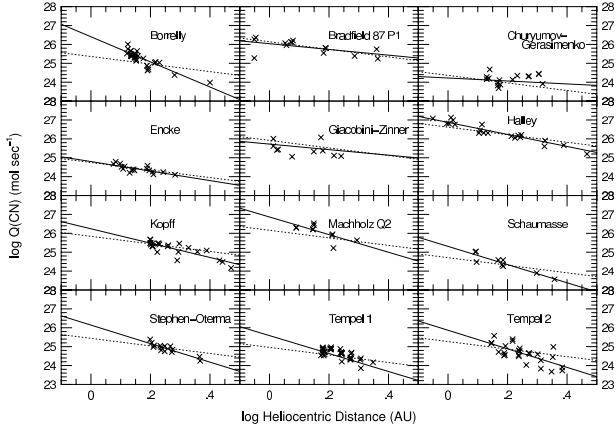}
\caption[fig11]{Twelve comets were observed at a large enough
range of heliocentric distances that we could investigate how the CN
production rates scale with heliocentric distance.  We did not attempt
to differentiate between photometric and non-photometric conditions, nor
did we separate apparitions.  These panels show the data along with the
line that best fits the data (solid line).  Also shown is a 1/R$_h^2$
line (dotted line).  The derived average CN values for 1\.{\sc au} using
the fit and the 1/R$_h^2$ law are shown in Table~\ref{cndata}.
}
\label{cnscale}
\end{figure}

\begin{figure}
\includegraphics[scale=0.33,angle=270]{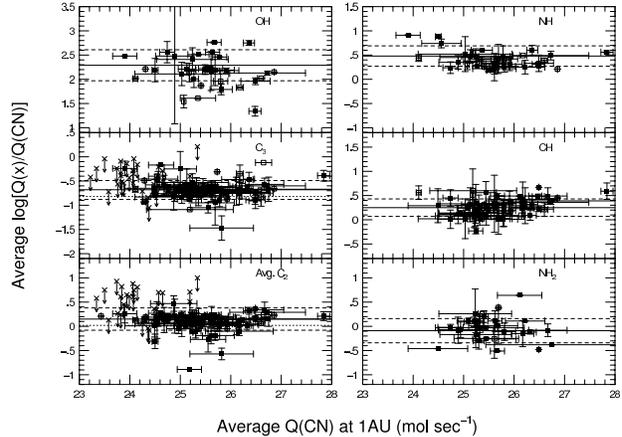}
\caption[fig12]{The average production rate ratios for each
comet and each molecule are shown plotted against the average Q(CN) at
1\,{\sc au}.  In each panel, the filled symbols are comets in the restricted
data set; the open symbols are the data not in the restricted data set.
For C$_{2}$ and C$_{3}$, the upper limits are shown for comets with detections
of CN but no C$_{2}$ or C$_{3}$ detection.  The solid horizontal line
is the weighted mean of the production rate ratios.  The dashed lines
represent the standard deviations from the means.  For C$_{2}$ and
C$_{3}$, the dotted lines mark the boundary between
typical and depleted comets. Details of this figure
are discussed in the text.
}
\label{avgwlim}
\end{figure}

For the discussion that follows, it is not critical if the CN varies
differently than we have assumed because
any deviation from this law will result in more scatter in the 1\,{\sc au}
values, and therefore larger error bars.  The exact values are not 
critical to our trends.  For most of our comets we did not have a large
enough range of heliocentric distance to investigate the correct scaling
with heliocentric distance; for twelve comets, we do cover large heliocentric
distance ranges and these are shown in Figure~\ref{cnscale}.
We fit a line to 
the data but did not in any way try to correct for weather or for 
changes from one apparition to the next.  The solid line shows the fit
while the dotted line shows a 1/R$_h^2$ line.  For some comets, there
is a big difference between the fit and the 1/R$_h^2$ line.
Table~\ref{cndata} lists the derived slopes and the averages from
using a 1/R$_h^2$ scaling versus the fitted scaling.  Not surprisingly,
we get different 1\,{\sc au} CN values depending on what we assume.
%\begin{landscape}
\begin{table*}%[ht!]
\small
\caption{The Heliocentric Dependence of the CN Production Rates}\label{cndata}
\vspace*{7pt}
\begin{tabular}{lrclccccc}
 & \multicolumn{3}{c}{Heliocentric} & & & Avg. Q(CN) & Avg. Q(CN) \\
 & \multicolumn{3}{c}{Distance} & & scaling & at 1\,{\sc {\sc au}} & at 1\,{\sc {\sc au}} \\
\multicolumn{1}{c}{Comet} & \multicolumn{3}{c}{Range ({\sc {\sc au}})} & Npts & slope & (fitted) & (1/R$_h^2$) \\
\hline
Borrelly & 1.32 & -- & 2.51 & 31 & -6.6 & $4.89\times10^{25}\pm3.43\times10^{25}$ & $2.53\times10^{26}\pm1.21\times10^{26}$ \\
Bradfield (87P1) & 0.89 & -- & 2.31 & 14 & -1.5 & $1.49\times10^{26}\pm9.05\times10^{24}$ & $1.30\times10^{26}\pm2.32\times10^{24}$ \\
CG & 1.35 & -- & 2.08 & 16 & -0.8 & $4.84\times10^{24}\pm3.38\times10^{24}$ & $2.57\times10^{24}\pm1.58\times10^{24}$ \\
Encke & 1.25 & -- & 1.92 & 17 & -2.5 & $5.50\times10^{24}\pm1.78\times10^{24}$ & $6.63\times10^{24}\pm2.05\times10^{24}$ \\
GZ & 1.03 & -- & 1.74 & 10 & -1.5 & $6.63\times10^{25}\pm7.41\times10^{25}$ & $5.71\times10^{25}\pm6.07\times10^{25}$ \\
Halley & 0.89 & -- & 4.84 & 22 & -3.2 & $4.67\times10^{26}\pm3.28\times10^{26}$ & $7.36\times10^{26}\pm2.84\times10^{26}$ \\
Kopff & 1.58 & -- & 2.77 & 24 & -3.8 & $6.39\times10^{25}\pm2.95\times10^{25}$ & $1.92\times10^{26}\pm8.52\times10^{25}$ \\
Machholz Q2 & 1.27 & -- & 1.97 & 10 & -4.7 & $3.09\times10^{26}\pm1.80\times10^{26}$ & $8.39\times10^{26}\pm4.28\times10^{26}$ \\
Schaumasse & 1.24 & -- & 2.28 & 9 & -4.8 & $7.71\times10^{24}\pm5.47\times10^{24}$ & $2.18\times10^{25}\pm7.82\times10^{24}$ \\
Stephan-Oterma & 1.58 & -- & 2.33 & 14 & -4.9 & $2.79\times10^{25}\pm1.27\times10^{25}$ & $1.45\times10^{26}\pm4.32\times10^{25}$ \\
Tempel 1 & 1.49 & -- & 2.39 & 43 & -4.0 & $1.36\times10^{25}\pm5.06\times10^{24}$ & $3.88\times10^{25}\pm1.31\times10^{25}$ \\
Tempel 2 & 1.40 & -- & 2.42 & 26 & -4.9 & $2.08\times10^{25}\pm2.01\times10^{25}$ & $1.07\times10^{26}\pm1.11\times10^{26}$ \\
\hline
\end{tabular}
\end{table*}
%\end{landscape}
However, as we only use the 1\,{\sc au} CN for Figure~\ref{avgwlim} (described
below) and that value does not affect our conclusions, the exact scaling
used is irrelevant, as it only moves points left or right and we care
most about vertical changes. None of the other species were scaled
to 1\,{\sc au}; all ratios were taken without scaling species to 1\,{\sc au}.

These average values are given in Table~\ref{averages}.
\begin{landscape}
\begin{table*}\tiny
\caption{Average Production Rate Ratios}\label{averages}
\vspace*{7pt}
\begin{tabular}{lc@{ }c@{  }c@{ }c@{  }c@{ }c@{  }c@{ }c@{  }c@{ }c@{  }c@{ }c@{  }c@{ }c}   %15 cols
\hline
\multicolumn{1}{c}{Comet} & log Q(CN) & & \multicolumn{12}{c}{Average log production rate ratios} \\ [3pt]
\cline{4-15} \\ [-4pt]
  &&    &&       &&       &&            &&       &  Average   \\
  & at 1\,{\sc au} & N$^a$ & OH/CN & N$^a$ & NH/CN & N$^a$ & C$_{3}$/CN & N$^a$ & CH/CN & N$^a$ & C$_{2}$/CN & N$^a$ & NH$_{2}$/CN & N$^a$  \\
\hline
{\bf 49P/Arend-Rigaux }
 &  24.56  $\pm$  0.47  &  7  &        &    &  0.74  $\pm$  0.085  &  2  &  -0.46  $\pm$  0.159  &  6  &        &    &  0.08  $\pm$  0.084  &  5  &        &    \\
{\bf C/1982 M1 (Austin) }
&  25.45  $\pm$  0.39  &  4  &        &    &        &    &  -0.63  $\pm$  0.175  &  4  &  0.08  $\pm$  0.190  &  2  &  0.19  $\pm$  0.029  &  3  &  -0.25  $\pm$  0.007  &  2  \\
{\bf C/1984 N1 (Austin) }
&  25.45  $\pm$  0.80  &  4  &        &    &        &    &  -0.65  $\pm$  0.104  &  3  &  0.23  $\pm$  0.100  &  3  &  0.18  $\pm$  0.083  &  3  &  -0.02      &  1  \\
{\bf C/1989 X1 (Austin) }
&  25.64  $\pm$  0.17  &  3  &  2.56      &  1  &  0.25  $\pm$  0.048  &  3  &  -0.70  $\pm$  0.020  &  3  &  0.31  $\pm$  0.048  &  3  &  0.22  $\pm$  0.045  &  3  &        &    \\
{\bf 19P/Borrelly }
&  25.69  $\pm$  0.47  &  31  &  2.19  $\pm$  0.318  &  10  &  0.42  $\pm$  0.297  &  14  &  -0.95  $\pm$  0.462  &  27  &  0.01  $\pm$  0.314  &  22  &  -0.19  $\pm$  0.096  &  29  &  0.01  $\pm$  0.362  &  11  \\
\hspace{1em}140P/Bowell-Skiff 
&  25.19  $\pm$  0.98  &  2  &        &    &        &    &  -1.09      &  1  &  0.60      &  1  &  0.19      &  1  &        &    \\
\hspace{1em}C/1979 Y1 (Bradfield) 
&  25.93      &  1  &        &    &        &    &  -0.87      &  1  &  0.18      &  1  &  -0.05      &  1  &        &    \\
{\bf C/1980 Y1 (Bradfield) }
&  26.49      &  1  &        &    &  0.27      &  1  &  -0.87      &  1  &  0.67      &  1  &  0.37      &  1  &  -0.48      &  1  \\
{\bf C/1987 P1 (Bradfield)  }
&  26.17  $\pm$  0.38  &  14  &        &    &        &    &  -0.74  $\pm$  0.115  &  14  &  0.26  $\pm$  0.124  &  14  &  0.09  $\pm$  0.069  &  14  &  -0.15  $\pm$  0.273  &  12  \\
\hspace{1em}16P/Brooks 2 
&  24.37  $\pm$  0.26  &  11  &        &    &        &    &  $<$-0.09      &    &        &    &  $<$0.61      &    &        &    \\
{\bf 23P/Brorsen-Metcalf }
&  25.67  $\pm$  0.28  &  3  &  2.76  $\pm$  0.021  &  2  &  0.25  $\pm$  0.110  &  3  &  -0.66  $\pm$  0.071  &  3  &  0.49  $\pm$  0.048  &  3  &  0.24  $\pm$  0.031  &  3  &  -0.51      &  1  \\
\hspace{1em}C/1983 O1 (Cernis) 
&  26.65  $\pm$  0.17  &  2  &        &    &        &    &  -0.12      &  1  &        &    &  0.24      &  1  &        &    \\
\hspace{1em}C/1990 E1 (CKN) 
&  25.32  $\pm$  0.24  &  2  &        &    &  0.41      &  1  &  -0.58  $\pm$  0.144  &  2  &  0.06  $\pm$  0.174  &  2  &  0.21  $\pm$  0.007  &  2  &        &    \\
\hspace{1em}C/1980 O1 (CP) 
&  24.63      &  1  &        &    &        &    &  $<$-0.16      &    &        &    &  $<$0.81      &    &        &    \\
{\bf 67P/CG }
&  24.68  $\pm$  0.31  &  16  &        &    &        &    &  -0.58  $\pm$  0.203  &  5  &        &    &  0.13  $\pm$  0.260  &  5  &        &  \\
{\bf 71P/Clark }
&  24.97  $\pm$  0.38  &  6  &        &    &        &    &  -0.71  $\pm$  0.129  &  2  &        &    &  -0.05  $\pm$  0.100  &  2  &        &  \\
{\bf 32P/Comas Sola }
&  25.18  $\pm$  0.29  &  2  &        &    &        &    &  -0.88      &  1  &        &    &  -0.89      &  1  &        &  \\
{\bf 27P/Crommelin }
&  25.23  $\pm$  0.65  &  8  &        &    &        &    &  -0.80  $\pm$  0.153  &  7  &  0.14  $\pm$  0.268  &  6  &  0.18  $\pm$  0.131  &  8  &  0.26  $\pm$  0.509  &  6\\
{\bf 6P/d'Arrest }
&  25.24  $\pm$  0.49  &  16  &  2.41  $\pm$  0.237  &  3  &  0.45  $\pm$  0.175  &  4  &  -0.53  $\pm$  0.178  &  16  &  0.33  $\pm$  0.405  &  11  &  0.21  $\pm$  0.074  &  15  &  -0.17  $\pm$  0.267  &  4\\
\hspace{1em}122P/de Vico 
&  26.60  $\pm$  0.08  &  2  &  2.02      &  1  &  0.39  $\pm$  0.050  &  2  &  -0.83  $\pm$  0.190  &  2  &  0.21  $\pm$  0.137  &  2  &  0.06  $\pm$  0.035  &  2  &        &  \\
\hspace{1em}79P/dTH 
&  23.60      &  1  &        &    &        &    &  $<$-0.24      &    &        &    &  $<$0.58      &    &        &  \\
{\bf 2P/Encke }
 &  24.74  $\pm$  0.23  &  17  &  2.56  $\pm$  0.225  &  8  &  0.24  $\pm$  0.121  &  8  &  -0.62  $\pm$  0.087  &  17  &  0.45  $\pm$  0.169  &  12  &  0.17  $\pm$  0.085  &  17  &        &  \\
\hspace{1em}4P/Faye 
&  24.43  $\pm$  0.16  &  2  &        &    &        &    &  $<$-0.52      &    &        &    &  $<$0.82      &    &        &  \\
\hspace{1em}37P/Forbes 
&  24.96  $\pm$  0.17  &  2  &        &    &        &    &  $<$-0.75      &    &        &    &  $<$0.34      &    &        &  \\
\hspace{1em}78P/Gehrels 2 
&  24.50  $\pm$  0.22  &  2  &        &    &        &    &  $<$-0.24      &    &        &    &  $<$0.93      &    &        &  \\
{\bf 21P/GZ }
&  25.82  $\pm$  0.91  &  10  &        &    &        &    &  -1.47  $\pm$  0.249  &  7  &        &    &  -0.57  $\pm$  0.122  &  9  &        &  \\
\hspace{1em}84P/Giclas 
&  24.83      &  1  &        &    &        &    &  $<$-0.71      &    &        &    &  $<$0.07      &    &        &  \\
{\bf 26P/GS }
&  24.49  $\pm$  0.47  &  7  &        &    &        &    &  -0.65  $\pm$  0.181  &  6  &  0.29  $\pm$  0.354  &  4  &  0.01  $\pm$  0.227  &  6  &  -0.46      &  1\\
\hspace{1em}65P/Gunn 
&  25.53  $\pm$  0.38  &  11  &        &    &        &    &  $<$-0.29      &    &        &    &  $<$0.69      &    &        &  \\
{\bf C/1995 O1 (Hale-Bopp) }
&  27.84  $\pm$  0.12  &  4  &        &    &  0.56  $\pm$  0.028  &  2  &  -0.39  $\pm$  0.114  &  3  &  0.59  $\pm$  0.167  &  2  &  0.21  $\pm$  0.086  &  3  &        &  \\
{\bf 1P/Halley }
&  26.67  $\pm$  0.51  &  25  &        &    &        &    &  -0.57  $\pm$  0.129  &  23  &  0.49  $\pm$  0.165  &  12  &  0.34  $\pm$  0.160  &  21  &  -0.08  $\pm$  0.128  &  10\\
\hspace{1em}51P/Harrington 
&  24.77      &  1  &        &    &        &    &  -0.74      &  1  &        &    &  0.17      &  1  &        &  \\
\hspace{1em}52P/Harrington-Abell 
&  24.51  $\pm$  0.25  &  3  &        &    &        &    &  $<$-0.43      &    &        &    &  $<$0.62      &    &        &  \\
\hspace{1em}C/1995 Q2 (HD) 
&  24.42  $\pm$  0.00  &  1  &        &    &        &    &  $<$-0.39      &    &        &    &  $<$0.24      &    &        &  \\
{\bf 1985 R1 (Hartley-Good) }
& 25.40  $\pm$  0.24  &  4  &        &    &        &    &  -0.68  $\pm$  0.033  &  4  &        &    &  0.11  $\pm$  0.018  &  4  &  0.20  $\pm$  0.125  &  4\\
\hspace{1em}161P/Hartley-IRAS 
&  25.37  $\pm$  0.48  &  2  &        &    &        &    &  -0.70      &  1  &        &    &  0.11      &  1  &  0.26      &  1\\
\hspace{1em}111P/HRC 
&  25.66      &  1  &        &    &        &    &  $<$-0.21      &    &        &    &  $<$0.70      &    &        &  \\
\hspace{1em}45P/HMP 
&  24.31      &  1  &  2.21      &  1  &        &    &  -0.93      &  1  &        &    &  0.19      &  1  &        &  \\
{\bf 88P/Howell }
&  25.00  $\pm$  0.31  &  3  &        &    &        &    &  -0.25  $\pm$  0.365  &  2  &        &    &  0.24      &  1  &        &  \\
\hspace{1em}C/1995 Y1 (Hyakutake)  
&  25.41      &  1  &  1.87      &  1  &  0.26      &  1  &  -0.80      &  1  &  0.31      &  1  &  0.16      &  1  &        &  \\
{\bf C/1996 B2 (Hyakutake)  }
&  26.73  $\pm$  0.68  &  3  &  2.13  $\pm$  0.035  &  2  &  0.49  $\pm$  0.087  &  3  &  -0.60  $\pm$  0.030  &  3  &  0.36  $\pm$  0.035  &  2  &  0.29  $\pm$  0.005  &  3  &        &  \\
{\bf 153P/Ikeya-Zhang }
&  26.48  $\pm$  0.14  &  3  &  1.34  $\pm$  0.102  &  3  &  0.31  $\pm$  0.035  &  3  &  -0.91  $\pm$  0.065  &  3  &  0.48  $\pm$  0.054  &  3  &  0.17  $\pm$  0.030  &  3  &        &  \\
{\bf C/1983 H1 (IAA) }
&  25.26  $\pm$  0.14  &  3  &        &    &        &    &  -0.66  $\pm$  0.025  &  3  &        &    &  0.04  $\pm$  0.052  &  3  &  0.12  $\pm$  0.014  &  2\\
\hspace{1em}126P/IRAS 
&  24.99  $\pm$  0.48  &  4  &        &    &        &    &  $<$-0.57      &    &        &    &  $<$0.36      &    &        &  \\
\hspace{1em}C/1983 O2 (IRAS) 
 &  24.85      &  1  &        &    &        &    &  $<$-0.29      &    &        &    &  $<$0.88      &    &        &  \\
{\bf 58P/Jackson-Neujmin }
&  23.90  $\pm$  0.27  &  5  &  2.48      &  1  &  0.91  $\pm$  0.014  &  2  &  -0.24  $\pm$  0.147  &  5  &        &    &  0.26  $\pm$  0.193  &  4  &        &  \\
{\bf 59P/Kearns-Kwee }
&  25.36  $\pm$  0.23  &  10  &  2.51  $\pm$  0.272  &  1  &  0.60  $\pm$  0.007  &  2  &  -0.77  $\pm$  0.241  &  8  &    &    &  0.14  $\pm$  0.230  &  8  &        &  \\
{\bf 68P/Klemola }
&  24.87  $\pm$  0.55  &  4  &        &    &        &    &  -0.81  $\pm$  0.290  &  3  &        &    &  0.47  $\pm$  0.164  &  3  &        &  \\
{\bf 75P/Kohoutek }
&  25.30  $\pm$  0.16  &  4  &        &    &        &    &  -0.67  $\pm$  0.127  &  4  &  0.25  $\pm$  0.260  &  4  &  -0.10  $\pm$  0.136  &  4  &  -0.30  $\pm$  0.172  &  4\\
\hline
\end{tabular}
\end{table*}
%\newpage
\begin{table*} \tiny
\begin{tabular}{lc@{ }c@{  }c@{ }c@{  }c@{ }c@{  }c@{ }c@{  }c@{ }c@{  }c@{ }c@{  }c@{ }c}   %15 cols
\hline
\multicolumn{1}{c}{Comet} & log Q(CN) & & \multicolumn{12}{c}{Average log production rate ratios} \\ [3pt]
\cline{4-15} \\ [-4pt]
  &&    &&       &&       &&            &&       &  Average   \\
  & at 1\,{\sc au} & N$^a$& OH/CN & N$^a$& NH/CN & N$^a$& C$_{3}$/CN & N$^a$& CH/CN & N$^a$& C$_{2}$/CN & N$^a$& NH$_{2}$/CN & N$^a$\\
\hline
{\bf 22P/Kopff }
&  25.81  $\pm$  0.30  &  24  &  1.79  $\pm$  0.113  &  3  &  0.27  $\pm$  0.124  &  3  &  -0.74  $\pm$  0.086  &  21  &  0.19  $\pm$  0.164  &  14  &  0.05  $\pm$  0.103  &  20  &  -0.11  $\pm$  0.178  &  9\\
{\bf 144P/Kushida }
&  25.27  $\pm$  0.16  &  4  &  2.01  $\pm$  0.177  &  4  &  0.28  $\pm$  0.014  &  2  &  -0.61  $\pm$  0.077  &  4  &  0.20  $\pm$  0.122  &  2  &  0.13  $\pm$  0.038  &  4  &        &  \\
{\bf C/1987 A1 (Levy) }
&  24.74  $\pm$  0.39  &  3  &        &    &        &    &  -0.61  $\pm$  0.018  &  3  &  0.01  $\pm$  0.190  &  2  &  0.14  $\pm$  0.071  &  3  &  -0.02  $\pm$  0.057  &  2\\
{\bf C/1990 K1 (Levy) }
&  26.36  $\pm$  0.20  &  5  &  2.75  $\pm$  0.055  &  3  &  0.60  $\pm$  0.098  &  4  &  -0.46  $\pm$  0.142  &  5  &  0.32  $\pm$  0.089  &  4  &  0.32  $\pm$  0.075  &  5  &        &  \\
{\bf C/1984 V1 (LR) }
&  25.21  $\pm$  0.24  &  6  &        &    &        &    &  -0.69  $\pm$  0.087  &  6  &  0.13  $\pm$  0.287  &  5  &  0.06  $\pm$  0.056  &  5  &  0.09  $\pm$  0.096  &  5\\
{\bf C/1988 A1 (Liller) }
&  26.32  $\pm$  0.22  &  5  &        &    &        &    &  -0.66  $\pm$  0.056  &  5  &  0.09  $\pm$  0.102  &  5  &  0.01  $\pm$  0.071  &  5  &  -0.12  $\pm$  0.043  &  5\\
\hspace{1em}C/2001 A2 (LINEAR) 
&  25.32      &  1  &        &    &        &    &  -0.68      &  1  &        &    &  0.27      &  1  &        &  \\
{\bf C/2000 WM1 (LINEAR) }
&  25.78  $\pm$  0.13  &  5  &  2.46      &  1  &  0.47  $\pm$  0.220  &  3  &  -0.89  $\pm$  0.136  &  5  &  0.02  $\pm$  0.159  &  2  &  0.02  $\pm$  0.117  &  5  &        &  \\
\hspace{1em}C/2006 VZ13 (LINEAR) 
&  25.35  $\pm$  0.35  &  2  &  1.61  $\pm$  0.014  &  2  &  0.37  $\pm$  0.093  &  2  &  -0.67  $\pm$  0.028  &  2  &  0.15      &  1  &  0.10  $\pm$  0.025  &  2  &        &  \\
\hspace{1em}93P/Lovas 1 
&  24.54  $\pm$  0.17  &  2  &        &    &        &    &  -0.40  $\pm$  0.042  &  2  &        &    &  0.13  $\pm$  0.375  &  2  &        &  \\
\hspace{1em}C/2007 E2 (Lovejoy) 
&  25.07      &  2  &  1.54  $\pm$  0.129  &  2  &  0.29  $\pm$  0.093  &  2  &  -0.78  $\pm$  0.035  &  2  &        &    &  0.07  $\pm$  0.024  &  2  &        &  \\
{\bf C/1994 T1 (Machholz) }
&  25.55  $\pm$  0.08  &  4  &  2.21  $\pm$  0.062  &  4  &  0.19      &  1  &  -0.62  $\pm$  0.080  &  3  &        &    &  0.21  $\pm$  0.042  &  4  &        &  \\
{\bf C/2004 Q2 (Machholz) }
&  26.49  $\pm$  0.34  &  10  &  1.96  $\pm$  0.067  &  9  &  0.33  $\pm$  0.174  &  9  &  -0.74  $\pm$  0.062  &  10  &  0.22  $\pm$  0.050  &  10  &  0.16  $\pm$  0.035  &  10  &        &  \\
\hspace{1em}141P/Machholz 2-A 
&  24.11  $\pm$  0.06  &  2  &  2.01  $\pm$  0.014  &  2  &  0.44  $\pm$  0.071  &  2  &  -0.67  $\pm$  0.064  &  2  &  0.56  $\pm$  0.144  &  2  &  0.13  $\pm$  0.019  &  2  &        &  \\
\hspace{1em}115P/Maury 
&  25.74      &  1  &        &    &        &    &  $<$-0.56      &    &        &    &  $<$0.27      &    &        &  \\
{\bf C/1993 Y1 (MR) }
&  25.03  $\pm$  0.61  &  7  &  2.11  $\pm$  0.241  &  5  &  0.52  $\pm$  0.359  &  7  &  -0.83  $\pm$  0.129  &  7  &  0.02  $\pm$  0.414  &  6  &  0.05  $\pm$  0.264  &  7  &        &  \\
{\bf C/1980 V1 (Meier) }
&  26.22  $\pm$  0.47  &  6  &        &    &  0.25  $\pm$  0.137  &  2  &  -0.71  $\pm$  0.165  &  6  &  0.29  $\pm$  0.371  &  5  &  0.14  $\pm$  0.020  &  5  &  0.11      &  1\\
\hspace{1em}C/1993 A1 (Mueller) 
&  26.18  $\pm$  0.61  &  2  &  1.83      &  1  &  0.49      &  1  &  -0.35  $\pm$  0.182  &  2  &  0.35      &  1  &  0.12      &  1  &        &  \\
{\bf 28P/Neujmin 1 }
&  24.62  $\pm$  0.25  &  2  &        &    &        &    &  -0.17      &  1  &        &    &  0.27      &  1  &        &  \\
{\bf C/1987 B1 (NTT) }
&  25.62  $\pm$  0.31  &  3  &        &    &        &    &  -0.63  $\pm$  0.135  &  3  &  0.25  $\pm$  0.050  &  2  &  0.23  $\pm$  0.176  &  3  &  0.11  $\pm$  0.205  &  2\\
\hspace{1em}C/1992 W1 (Oshita) 
&  23.43      &  1  &        &    &        &    &  $<$-0.42      &    &        &    &  0.21      &  1  &        &  \\
{\bf C/1980 Y2 (Panther) }
&  26.49  $\pm$  0.24  &  4  &        &    &        &    &  -0.89  $\pm$  0.181  &  3  &        &    &  0.13  $\pm$  0.077  &  3  &        &  \\
{\bf 80P/Peters-Hartley }
&  25.24  $\pm$  0.14  &  2  &        &    &        &    &  -0.76  $\pm$  0.122  &  2  &  -0.23  $\pm$  0.057  &  2  &  0.15  $\pm$  0.189  &  2  &  0.01  $\pm$  0.144  &  2\\
\hspace{1em}83P/Russell 1 
&  26.13  $\pm$  0.36  &  3  &        &    &        &    &  $<$-0.40      &    &        &    &  $<$0.71      &    &        &  \\
\hspace{1em}91P/Russell 3 
&  24.53      &  1  &        &    &        &    &  $<$-0.04      &    &        &    &  $<$0.75      &    &        &  \\
{\bf 24P/Schaumasse }
&  24.89  $\pm$  0.77  &  9  &  2.48  $\pm$  1.398  &  4  &  0.35  $\pm$  0.136  &  3  &  -0.76  $\pm$  0.159  &  8  &  0.11  $\pm$  0.195  &  3  &  0.21  $\pm$  0.349  &  8  &  -0.08  $\pm$  0.272  &  2\\
\hspace{1em}106P/Schuster 
&  23.99      &  1  &        &    &        &    &  $<$-0.22      &    &        &    &  $<$0.61      &    &        &  \\
{\bf 31P/SW2 }
&  25.61  $\pm$  0.16  &  6  &  2.17  $\pm$  0.397  &  2  &  0.32  $\pm$  0.205  &  2  &  -0.75  $\pm$  0.174  &  2  &        &    &  -0.20  $\pm$  0.156  &  2  &        &  \\
{\bf 102P/Shoemaker 1 }
&  26.11  $\pm$  0.45  &  3  &        &    &        &    &  -0.71  $\pm$  0.288  &  3  &  0.49      &  1  &  0.03  $\pm$  0.204  &  3  &  0.64      &  1\\
\hspace{1em}C/1984 K1 (Shoemaker) 
&  26.13  $\pm$  0.35  &  2  &        &    &        &    &  -0.47      &  1  &        &    &  $<$0.51      &    &        &  \\
\hspace{1em}C/1984 U2 (Shoemaker) 
&  24.58  $\pm$  0.21  &  2  &        &    &        &    &  $<$-0.84      &    &        &    &  $<$-0.04      &    &        &  \\
\hspace{1em}C/1987 H1 (Shoemaker) 
&  26.85      &  1  &        &    &        &    &  $<$0.21      &    &        &    &  $<$1.00      &    &        &  \\
\hspace{1em}C/1989 A6 (Shoemaker) 
&  25.73      &  1  &        &    &        &    &  -0.31      &  1  &        &    &  -0.05      &  1  &        &  \\
{\bf C/1988 J1 (SH) }
&  25.25      &  1  &        &    &        &    &  -0.57      &  1  &        &    &  -0.12      &  1  &        &  \\
\hspace{1em}192P/Shoemaker-Levy 1  
&  24.11      &  1  &        &    &        &    &  $<$-0.72      &    &        &    &  $<$0.13      &    &        &  \\
{\bf C/2002 E2 (SM) }
&  25.49  $\pm$  0.04  &  3  &  2.20  $\pm$  0.097  &  3  &  0.19  $\pm$  0.032  &  3  &  -0.76  $\pm$  0.050  &  3  &  0.21  $\pm$  0.021  &  3  &  0.20  $\pm$  0.015  &  3  &        &  \\
{\bf C/1986 V1 (Sorrells) }
&  26.15  $\pm$  0.21  &  3  &        &    &        &    &  -0.69  $\pm$  0.035  &  3  &  0.23  $\pm$  0.221  &  2  &  0.02  $\pm$  0.062  &  3  &        &  \\
{\bf 38P/Stephan-Oterma }
&  25.44  $\pm$  0.30  &  14  &        &    &        &    &  -0.69  $\pm$  0.133  &  12  &  0.56  $\pm$  0.231  &  5  &  0.03  $\pm$  0.074  &  11  &        &  \\
\hspace{1em}C/1983 J1 (SSF) 
&  24.29  $\pm$  0.01  &  2  &        &    &        &    &  -0.92  $\pm$  0.035  &  2  &        &    &  0.14  $\pm$  0.057  &  2  &        &  \\
{\bf 64P/Swift-Gehrels }
&  24.94  $\pm$  0.50  &  10  &        &    &        &    &  -0.68  $\pm$  0.057  &  10  &  0.13  $\pm$  0.064  &  2  &  0.14  $\pm$  0.080  &  8  &  -0.07  $\pm$  0.158  &  3\\
\hspace{1em}109P/Swift-Tuttle 
&  26.86      &  1  &  2.15      &  1  &  0.21      &  1  &  -0.66      &  1  &  0.45      &  1  &  0.23      &  1  &        &  \\
{\bf C/1996 B1 (Szczepanski) }
&  25.91  $\pm$  0.24  &  3  &  2.18  $\pm$  0.053  &  3  &  0.31  $\pm$  0.127  &  3  &  -0.75  $\pm$  0.039  &  3  &  0.35  $\pm$  0.243  &  3  &  0.07  $\pm$  0.015  &  3  &        &  \\
\hspace{1em}98P/Takamizawa 
&  24.54      &  1  &        &    &        &    &  $<$-0.14      &    &        &    &  $<$0.65      &    &        &  \\
\hspace{1em}C/1994 J2 (Takamizawa) 
&  25.80  $\pm$  0.14  &  2  &  1.96  $\pm$  0.230  &  2  &  0.39  $\pm$  0.050  &  2  &  -0.69  $\pm$  0.035  &  2  &  0.13  $\pm$  0.272  &  2  &  0.08  $\pm$  0.085  &  2  &        &  \\
{\bf C/1994 G1-A (TL) }
&  25.60  $\pm$  0.24  &  3  &  2.22  $\pm$  0.045  &  3  &  0.33  $\pm$  0.049  &  3  &  -0.69  $\pm$  0.023  &  3  &  0.29  $\pm$  0.099  &  3  &  0.16  $\pm$  0.012  &  3  &        &  \\
\hspace{1em}69P/Taylor 
&  24.67  $\pm$  0.34  &  5  &        &    &        &    &  $<$-0.35      &    &        &    &  $<$0.60      &    &        &  \\
{\bf 9P/Tempel 1 }
&  25.13  $\pm$  0.21  &  43  &  2.21  $\pm$  0.124  &  31  &  0.47  $\pm$  0.144  &  28  &  -0.58  $\pm$  0.334  &  40  &  0.31  $\pm$  0.346  &  31  &  0.07  $\pm$  0.092  &  39  &        &  \\
{\bf 10P/Tempel 2 }
&  25.32  $\pm$  0.58  &  26  &        &    &        &    &  -0.63  $\pm$  0.197  &  18  &  0.08  $\pm$  0.100  &  9  &  0.05  $\pm$  0.152  &  19  &  -0.03  $\pm$  0.282  &  8\\
\hspace{1em}55P/Tempel-Tuttle
&  25.18  $\pm$  0.02  &  2  &  2.14  $\pm$  0.190  &  2  &  0.55  $\pm$  0.050  &  2  &  -0.64  $\pm$  0.057  &  2  &  0.51  $\pm$  0.540  &  2  &  0.12  $\pm$  0.014  &  2  &        &  \\
\hline
\end{tabular}
\end{table*}
%\newpage
\begin{table*} \tiny
\begin{tabular}{lc@{ }c@{  }c@{ }c@{  }c@{ }c@{  }c@{ }c@{  }c@{ }c@{  }c@{ }c@{  }c@{ }c}   %15 cols
\hline
\multicolumn{1}{c}{Comet} & log Q(CN) & & \multicolumn{12}{c}{Average log production rate ratios} \\ [3pt]
\cline{4-15} \\ [-4pt]
  &&    &&       &&       &&            &&       &  Average   \\
  & at 1\,{\sc au} & N$^a$& OH/CN & N$^a$& NH/CN & N$^a$& C$_{3}$/CN & N$^a$& CH/CN & N$^a$& C$_{2}$/CN & N$^a$& NH$_{2}$/CN & N$^a$\\
\hline\\
\hspace{1em}C/1987 B2 (Terasako) 
&  25.37      &  1  &        &    &        &    &  -0.78      &  1  &  0.02      &  1  &  -0.04      &  1  &        &  \\
\hspace{1em}C1985 T1 (Thiele) 
&  25.69      &  1  &        &    &        &    &  -0.58      &  1  &        &    &  0.10      &  1  &  0.39      &  1\\
{\bf 62P/Tsuchinshan 1 }
&  25.10  $\pm$  0.25  &  6  &        &    &        &    &  -0.70  $\pm$  0.072  &  6  &  0.24  $\pm$  0.170  &  3  &  0.02  $\pm$  0.066  &  6  &  0.11  $\pm$  0.057  &  2\\
{\bf 8P/Tuttle }
&  25.61  $\pm$  0.56  &  2  &        &    &  0.22  $\pm$  0.246  &  2  &  -0.62  $\pm$  0.122  &  2  &  0.33  $\pm$  0.592  &  2  &  0.07  $\pm$  0.085  &  2  &  -0.26  $\pm$  0.397  &  2\\
\hspace{1em}41P/TGK 
&  24.29      &  1  &        &    &        &    &  -0.49      &  1  &        &    &  0.15      &  1  &        &  \\
\hspace{1em}40P/Vaisala 
&  24.28      &  1  &        &    &        &    &  $<$-0.56      &    &        &    &  $<$0.27      &    &        &  \\
{\bf 76P/WKI }
&  24.50  $\pm$  0.06  &  2  &  2.19  $\pm$  0.238  &  2  &  0.88  $\pm$  0.007  &  2  &  -0.75      &  1  &        &    &  -0.31  $\pm$  0.148  &  2  &        &  \\
{\bf 81P/Wild 2 }
&  25.55  $\pm$  0.44  &  7  &        &    &        &    &  -1.05  $\pm$  0.114  &  2  &        &    &  -0.27  $\pm$  0.257  &  2  &        &  \\
{\bf C/1986 P1 (Wilson) }
&  26.75  $\pm$  0.51  &  4  &        &    &        &    &  -0.67  $\pm$  0.251  &  3  &  0.39      &  1  &  0.15  $\pm$  0.150  &  3  &  -0.38      &  1\\
\hspace{1em}114P/Wiseman-Skiff 
&  24.34  $\pm$  0.32  &  3  &        &    &        &    &  $<$-0.54      &    &        &    &  $<$0.45      &    &        &  \\
\hspace{1em}43P/Wolf-Harrington 
&  24.97  $\pm$  0.16  &  2  &        &    &        &    &  $<$-1.05      &    &        &    &  $<$-0.08      &    &        &  \\
{\bf C/1989 A1 (Yanaka) }
 &  26.17  $\pm$  0.57  &  4  &        &    &        &    &  -0.75  $\pm$  0.235  &  4  &  0.42  $\pm$  0.356  &  3  &  -0.10  $\pm$  0.093  &  2  &        &  \\
\hline
\\ [-4pt]
\multicolumn{15}{l}{$a$: N = number of points used in average} \\
\multicolumn{15}{l}{Comet names in boldface are in restricted data set; indented names are not in restricted set} \\ [3pt]
\underline{Comet Abbreviations:} \\ [3pt]
\multicolumn{5}{p{2.0in}}{\hspace{1em}CKN = Cernis-Kiuchi-Nakamura} &
\multicolumn{5}{p{2.0in}}{HRC = Helin-Roman-Crockett} &
\multicolumn{5}{p{2.0in}}{NTT = Nishikawa-Takamizawa-Tago} \\

\multicolumn{5}{p{2.0in}}{\hspace{1em}CP = Cernis-Petrauskas} &
\multicolumn{5}{p{2.0in}}{HMP = Honda-Mrkos-Pajdusokova} &
\multicolumn{5}{p{2.0in}}{SH = Shoemaker-Holt} \\

\multicolumn{5}{p{2.0in}}{\hspace{1em}CG = Churyumov-Gerasimenko} &
\multicolumn{5}{p{2.0in}}{IAA = IRAS-Araki-Alcock} &
\multicolumn{5}{p{2.0in}}{SSF = Sugano-Saigusa-Fujikawa} \\

\multicolumn{5}{p{2.0in}}{\hspace{1em}dTH = Du~Toit-Hartley} &
\multicolumn{5}{p{2.0in}}{LR = Levy-Rudenko} &
\multicolumn{5}{p{2.0in}}{TL = Takamizawa-Levy} \\

\multicolumn{5}{p{2.0in}}{\hspace{1em}GZ = Giacobini-Zinner} &
\multicolumn{5}{p{2.0in}}{Machholz 2A = Machholz Piece 2A} &
\multicolumn{5}{p{2.0in}}{TGK = Tuttle-Giacobini-Kresak} \\

\multicolumn{5}{p{2.0in}}{\hspace{1em}GS = Grigg-Skjellerup} &
\multicolumn{5}{p{2.0in}}{MR = McNaught-Russell} &
\multicolumn{5}{p{2.0in}}{WKI = West-Kohoutek-Ikemura} \\

\multicolumn{5}{p{2.0in}}{\hspace{1em}HD = Hartley-Drinkwater} \\
\hline
\end{tabular}

\end{table*}
\end{landscape}
In Table~\ref{averages} we include the comet name (sometimes
abbreviated to save space; the abbreviations are at the
bottom of the table), the average CN production rate at 1\,{\sc au}
and the average log[Q(X)/Q(CN)] where X represents
the various observed species.  We also include the standard deviations
of the averages (when more than one data point was present). 
The number of data points going into each average is listed.
When we did not observe a species (either because
it was not in our bandpass or we could not measure it) the value in
the table is either the upper limit (for C$_{2}$ or C$_{3}$ as described
above) or is left blank.  

As noted above, a number of comets were observed on only one or two
nights.  While these observations may have been of extremely high quality
(comet 122/de~Vico is such an example), many of the single or double
night observations were also of faint and less well observed comets.  Thus,
in a manner similar to AH95, we created a restricted
data set for exploration of the data. For our purposes, we define
a comet as being a member of the restricted data set if it was observed
on at least three different nights (or night/position angle combinations).
There are 72 comets that meet this criterion.  However, of these 72
comets, 13 had only upper limits for C$_{2}$ and/or C$_{3}$.
These upper limits are not constraining and so we dropped these 13 comets
from our definition of the restricted data set.  Thus, our restricted
data set consists of the remaining 59 comets.
In Table~\ref{averages}
these comets are listed with their names left-justified and bold-faced.
The 18 single-observation and 20 double-observation comets are indented
and not bold-faced, as are the 13 comets with more than 2 observations
but only upper limits for C$_{3}$ and/or C$_{2}$.
This definition of restricted data set is somewhat
arbitrary and does not attempt to take into account the quality of
the spectra.  However, it is an unbiased approach to choosing the restricted
set.

Figure~\ref{avgwlim} shows the production rate ratios of all comets as 
a function of the 1\,{\sc au} CN production rate.  
While it would have been preferable to plot against CN/OH, we did not
observe OH on enough dates to use the CN/OH data in the comparison.
AH95 showed that CN correlates well with OH (their figure 9) for
all types of comets.
The data from the restricted data set are shown as filled symbols, while
the data that are not in the restricted data set are shown with open
symbols
The upper limits
for C$_{3}$ and C$_{2}$ discussed above are shown on the appropriate panels.
As indicated above, almost none of these limits is very constraining.  When
we did not detect a species, it did not necessarily mean that there was
none of that molecule, just that the noise was high. 
Included in Figure~\ref{avgwlim} are the weighted means and standard
deviations for
the detections in the restricted data set for the quantities on the y axes.  
These means and standard deviations were computed using all 59 of the
comets in the restricted data set, without consideration of extreme
outliers.

The scatter in the Q(OH)/Q(CN) is quite high (confirmed by the standard
deviation being the largest we have) and it is difficult to reach
any conclusions based on these data.  This is another reason why we
chose not to compare all the comets with ratios to OH production.  
For the comets in the restricted data set, the outliers are 23P/Brorsen-Metcalf
and C/1990~K1 (Levy) on the high side and 22P/Kopff and 153P/Ikeya-Zhang
on the low side.  It is Ikeya-Zhang that is the extremely low point.

Q(NH)/Q(CN) is restricted to a very narrow range.
The three high data points are
49P/Arend-Rigaux, 58P/Jackson-Neujmin and 76P/West-Kohoutek-Ikemura.
These are of sufficiently good quality that they actually move the weighted
mean upward over what the eye would choose.  There are a number of comets
below the bottom standard deviation line. C/1994~T1 (Machholz) and C/2002~E2 
(Snyder-Murakami) are the most deviant.  Five other comets are within 0.05
of the lower standard deviation.

Q(NH$_{2}$)/Q(CN) shows more scatter than Q(NH)/Q(CN), though one would assume
that NH$_{2}$ dissociates to NH.  Generally, we only observed the NH$_{2}$ 
with the IDS (with a few exceptions) while we observed the NH with the LCS.
Rarely are the two species observed in the same comet due to the wavelength
range covered in the spectra.  
Inspection of Table~\ref{averages} shows that the error bars are
generally quite a bit higher for the detections of NH$_{2}$.
The highest Q(NH$_{2}$)/Q(CN)
is for comet 102P/Shoemaker 1, which had only one NH$_{2}$ detection when
the comet was at 2.01\,{\sc au} and no NH measurement. 
27P/Crommelin and 1985~R1 (Hartley-Good) are slightly elevated in NH$_{2}$.
The lowest outliers are 23P/Brorsen-Metcalf,
C/1980~Y1 (Bradfield), 26P/Grigg-Skjellerup and C/1986~P1 (Wilson).  These
comets had only one NH$_{2}$ detection each.  NH$_{2}$ is a very difficult
feature to observe since it is very weak and diffuse.
Comets Brorsen-Metcalf and Bradfield both also had NH observations.  Both
were at the lower edge or just below the NH normal band.

Q(CH)/Q(CN) seems to show many comets that are off the average by
just about the standard deviation of the average.  However, there are
few extreme outliers.  Most of the extreme comets, such as 
C/1980~Y1 (Bradfield), with log[Q(CH)/Q(CN)]=0.67, have only 1 data point.
However, 2P/Encke, 1P/Halley, C/1993~Y1 (McNaught-Russell), and 
38P/Stephan-Oterma have many data points.
All of these comets are at the high end, except for McNaught-Russell.
23P/Brorsen-Metcalf has three data points and is also high.
The lowest data point is comet 80P/Peters-Hartley, with two data points when
the comet was at 1.78\,{\sc au}.
  
If we ignore the upper limits, Q(C$_{2}$)/Q(CN) shows a very tight 
correlation for most comets but some comets show significant
deviations.  The high data point is 68P/Klemola.  The low
outliers are comets 19P/Borrelly, 32P/Comas Sola,
21P/Giacobini-Zinner,
75P/Kohoutek,
C/1988~J1 (Shoemaker-Holt),
31P/Schwassmann-Wachmann~2,
76P/West-Kohoutek-Ikemura,
81P/Wild 2,
and C/1989~A1 (Yanaka).  Most of
these are well observed.  The two lowest comets are Giacobini-Zinner (-0.57)
and Comas Sola (-0.89).

Q(C$_{3}$)/Q(CN) also shows high scatter, though most of the upper
limits are not very constraining.  The high outliers are 49P/Arend-Rigaux,
C/1995~O1 (Hale-Bopp), 88P/Howell, 58P/Jackson-Neujmin, C/1990~K1 (Levy),
and 28P/Neujmin~1.  Neujmin~1, Jackson-Neujmin and Howell are the three
highest values.  On the low side, 21P/Giacobini-Zinner is the extremely
low point, with log[Q(C$_{3}$)/Q(CN)]=-1.47.  The other low points (values
less than -0.9) are 
19P/Borrelly, 153P/Ikeya-Zhang, and 81P/Wild~2.

Categorizing individual comets was not the point of our program.  We sought
to understand whether comets could have been formed with more than
one compositional type.  This is not the first study of its kind.
In addition to the papers by AH95, F09 and
LS11 listed in the introduction, smaller
studies were carried out by A'Hearn and Millis (1980), Newburn and
Spinrad (1984), Fink and Hicks (1996) and by our group
(Cochran 1987; Cochran {\it et al.} 1987, 1989, 1992; Cochran and Barker
1987).  \nocite{fihi96,ahmi80,nesp84}
Thus, it has already been discovered that, while the spectra of most
comets are very similar, some comets show much weaker C$_{2}$ and C$_{3}$
relative to other species.  AH95 used a database
of 85 comets to quantify this trend and have defined two taxonomic
classes: typical and carbon-chain depleted (shortened to depleted).

In order to study the abundance correlations of the various species,
we started by forming weighted means and standard deviations of
the data.  These are listed in Table~\ref{ratios}.
\begin{table*}[ht!]
\footnotesize
\caption{Derived Production Rate Ratios}\label{ratios}
\vspace*{7pt}
\begin{tabular}{lcccccc}   % 7 cols
\hline
 & \multicolumn{6}{c}{Log Production Rate Ratio with respect to CN} \\
\cline{2-7}
 & OH & NH & C$_{3}$ & CH & C$_{2}$ & NH$_{2}$  \\
 &    &    &         &    & (average)          \\
\hline
All Comets & $2.09 \pm 0.35$ & $0.46 \pm 0.19$ & $-0.68 \pm 0.20$ &
  $0.25 \pm 0.19$ & $0.15 \pm 0.20$ & $-0.09 \pm 0.27$  \\
\hspace*{1em}(110 comets) & (33 comets) & (38 comets) & (84 comets) & (41 comets) & (84 comets)& (29 comets) \\ [5pt]
All Restricted Comets & $2.29 \pm 0.32$ & $0.48 \pm 0.21$ & $-0.68 \pm 0.19$ &
  $0.25 \pm 0.18$ & $0.15 \pm 0.23$ & $-0.09 \pm 0.25 $  \\
\hspace*{1em}(59 comets) & (23 comets) & (27 comets) & (59 comets) & (41 comets) & (59 comets)& (26 comets) \\ [5pt]
\hline
Restricted, Typical & $2.29 \pm 0.33$ & $0.48 \pm 0.21$ & $-0.67 \pm 0.15$ &
  $0.25 \pm 0.18$ & $0.16 \pm 0.14$ & $-0.09 \pm 0.26 $  \\
\hspace*{1em}(54 comets) & (21 comets) & (25 comets) & (54 comets) & (39 comets) & (54 comets)& (25 comets)  \\ [5pt]
Restricted, Depleted & $2.19 \pm 0.16$ & $0.44 \pm 0.28$ & $-1.06 \pm 0.39$ &
  $0.02 \pm 0.26$ & $-0.24 \pm 0.60$ & 0.01   \\
\hspace*{1em}(5 comets) & (2 comets) & (2 comets) & (5 comets) & (2 comets) & (5 comets)& (1 comet) \\ [5pt]
\hline
\multicolumn{7}{c}{Depleted comet is defined as log [Q(C$_{3}$)/Q(CN)] $\leq$ -0.86 and log [Q(C$_{2}$)/Q(CN)] $\leq$ 0.02.}
\end{tabular}
\end{table*}
In this table, we list the values for each species over various
subsets of the data.  First, we computed the mean for all 110 comets
with CN observations (upper limits were not used in the means).  We
also list the means and standard deviations for the complete
restricted data set.  Inspection of Table~\ref{ratios} shows
that the restricted data set is
representative of the whole data set. The complete restricted data
set's means and standard deviations are shown as solid and dashed
lines, respectively, in Figure~\ref{avgwlim}.

Using the values from
the complete data set, we have formed a definition for a depleted
comet in our data set.  A depleted comet is one for which
$log [Q(C_{3})/Q(CN)] \leq -0.86$ \underline{and}
$\log [Q(C_{2})/Q(CN)] \leq 0.02$.  Note that we require 
both the C$_{3}$ and C$_{2}$ ratios to be low.  These values
are shown in Figure~\ref{avgwlim} as dotted lines in the C$_{2}$ and
C$_{3}$ panels. Within the restricted
data set we find 5 comets that fit this definition: 19P/Borrelly,
32P/Comas Sola, 21P/Giacobini-Zinner, C/2000~WM1 (LINEAR), and 81P/Wild~2.
Borrelly, Giacobini-Zinner and Wild 2 have been previously shown
to be depleted by us and others. 
Thus, in our restricted data set we find 5/59 comets belong
to the carbon-chain depleted group, or 9\%.
For the comets not in the restricted set,
C/1979~Y1 (Bradfield) is also depleted. In addition, while most of
the upper limits are not constraining, two of the comets with only
upper limits have values that are below the defined cutoff for
depleted comets and therefore must be depleted:
C/1984~U2 (Shoemaker) and 43P/Wolf-Harrington (both were found depleted
by AH95).

\begin{figure}[hb!]
\includegraphics[scale=0.33,angle=270]{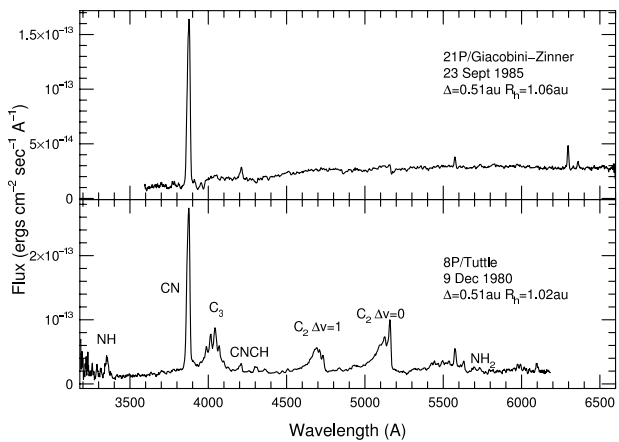}
\caption[fig13]{IDS Spectra of 21P/Giacobini-Zinner and 8P/Tuttle are
shown when both comets were near a heliocentric distance of 1\,{\sc au} and
a geocentric distance of 0.5\,{\sc {\sc au}}.  While both comets show
similar CN (including the weaker $\Delta v=-1$ band near $\sim$4200\,\AA),
there is little evidence of the C$_{3}$ or C$_{2}$ bands in the Giacobini-Zinner
spectrum.
}
\label{gz}
\end{figure}

For these depleted comets, the depletion of C$_{2}$ and C$_{3}$ is not a
subtle effect.  Figure~\ref{gz} shows spectra of 21P/Giacobini-Zinner
and 8P/Tuttle, obtained with the same instrument. 
The viewing geometry for these two observations is almost identical, with
only a very slight difference in heliocentric distance.  The two spectra
are scaled by the CN band.  The Tuttle spectrum shows all of the usual
molecules while the Giacobini-Zinner spectrum is ``missing" the C$_{2}$ and
C$_{3}$.  Inspection of the Giacobini-Zinner spectrum shows we
detected the weak CN $\Delta v = -1$ band at $\sim4200$\,\AA, so the
failure to detect the C$_{2}$ and C$_{3}$ is not a failure of signal/noise.
Giacobini-Zinner really does have very little of these two molecules!

In Table~\ref{ratios} we list means and standard deviations for
the 54 comets in the restricted data set that are not depleted (``typical")
and separately list these values for the 5 depleted comets.  Comparison
of these two groups shows that, while the depleted comets are significantly
different in their C$_{3}$ and C$_{2}$ ratios, OH and NH look the
same in both sets.  CH shows a little difference between the typical
and depleted comets, but the error bars are high.  NH$_{2}$ has
too little data (only 1 comet) in the depleted group to draw any conclusion.

As noted above, other groups have looked at abundance correlations
between comets.
As each study has used slightly different fluorescence efficiencies and
Haser scale lengths, one cannot simply 
compare the production rate ratios from one study with another
study.  However, as AH95 pointed out 
``...the pattern of values of the ratios will, to first order, be 
independent of these parameters".

\begin{figure}[ht!]
\includegraphics[scale=0.45]{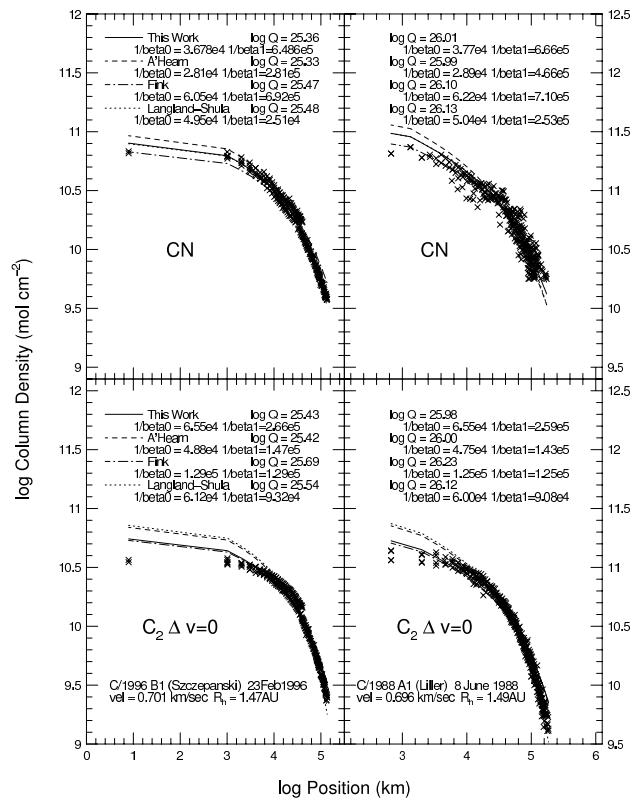}
\caption[fig14]{The CN and C$_{2}$ data for comets C/1988~A1
(Liller)
and C/1996~B1 (Szczepanski) are shown along with Haser model fits using
the scale lengths of this paper, AH95, F09 and LS11.  The two comets were
observed at comparable heliocentric distances.  For C$_{2}$, the fits using our
scale lengths are comparable to those using the scale lengths of F09.
The AH95 and LS11 fits are similar.  For CN, our fits are indistinguishable
from ones using the scale lengths of LS11.  The AH95 and
F09 models are higher and lower, respectively.  However, note the different
production rates that are computed with these fits.  See text for discussion.
}
\label{c2_scale}
\end{figure}

Figure~\ref{c2_scale} illustrates the effects of using different scale
lengths for computing the column densities.  It shows the CN and C$_{2}$ column
densities for two different comets along with model fits using the
scale lengths of AH95, F09, LS11 and  the values used in this paper.
These two particular cometary data sets were chosen because
they are LCS data with lots of points, they are high quality so
that the trends are well defined, there is very little asymmetry seen and
they are both at approximately the same heliocentric distance.  This
distance is sufficiently far from 1\,{\sc au} that the various heliocentric
scaling laws play a significant role in the scale lengths.  
C$_{2}$ was chosen because it is in the bandpass of all of the surveys.
CN is the molecule used in the denominator of the ratios.
Inspection of the two C$_{2}$ panels shows that our scale lengths do a
reasonably
good job of fitting the comet Liller data, though they may underestimate
the column densities around 10$^4$ km from the optocenter.  For
comet Szczepanski, the fit is good at cometocentric distances greater
than 1000\,km but the inner part of the coma is fit poorly.  This illustrates
that the Haser model does not handle the two-step (grandparent to parent to
daughter) dissociation well. The models using the F09 scale lengths are
almost indistinguishable in shape to those with our scale lengths,
but the derived log production rates differ from ours by $\sim$0.25.
The model fits using the AH95 and LS11 scale lengths are very similar.
However, the production rates for the AH95
scale lengths are essentially the same as ours while the LS11 production
rates are intermediate to ours and F09. 
For CN, our model fits and those
of LS11 are indistinguishable but yield production rates that differ
by 0.12.  Models using F09 and AH95 scale lengths are below and
above our model fits, respectively.  Again, our production rates
match those derived from AH95 scale lengths, while LS11 and F09 scale lengths
yield the same, higher, production rates. For Szczepanski, our CN scale lengths
fit extremely well, while we overestimate the column densities in the inner
coma of Liller.

The scale lengths used
for each model are on each panel and are considerably different
for the different groups.  It is a well known effect that there is a
family of Haser model scale lengths allowed for each molecule
since the parent and daughter
scale lengths are degenerate in the model and can compensate for one another. 
There
is nothing in Figure~\ref{c2_scale} to warrant picking another set
of scale lengths over the ones we have used in this paper
(Table~\ref{scalelengths}). What is a bit more critical is the boundary
for declaring a comet to be typical or depleted.  We would say that
Liller was slightly depleted.  AH95
listed Liller, though not in their restricted set, as typical.  Using their
scale lengths and our cutoff, we would still find Liller to be depleted.
With the AH95 definition and either our or their scale lengths Liller would
be typical.  We found Szczepanski to be typical, as did F09.

The recent study of LS11 has far fewer comets
than our study or the others we compare to.  However, the data should
be comparable to ours as they used a long slit CCD instrument with a 
bandpass similar to the LCS.  They suffered from difficult sky
subtraction because of street lamps near Lick Observatory, something
they took great pains to deal with.  McDonald does not
have this problem.  Also, some of
their observations were at twilight when the sky contribution was
more severe.  We almost never had to observe under such conditions
so did not encounter these problems.

F09 used a long slit CCD instrument
but concentrated on the red region of the spectrum. Thus, he was
unable to detect OH, NH or C$_{3}$.  For CN, he observed the
CN red system instead of the violet system we observed.
F09 observed NH$_{2}$ and O~($^1$D).  The latter could be used
to derive Q(H$_2$O).

AH95 used a fundamentally different data type.
Their survey used photometry and narrow band filters optimized
for comets.  With photometry, the apertures are considerably
larger than with our spectrograph and so AH95  are much
more sensitive at low signal levels.  They cannot, however, detect the very 
weak bands, such as CH.  In addition, since the different species
are observed sequentially and not simultaneously, they require
photometric conditions.  Analyses of photometric observations
must assume spherical symmetry
of the coma since they get little spatial information.

AH95 looked at 85 comets to determine abundance
patterns; 41 of those comets formed their restricted data set over which
they drew conclusions.  They found that 12/41 (29\%) of the comets were in the
depleted group.  This is significantly more than we found.  However,
AH96 used a definition based {\it only} on a low value for
C$_{2}$.  Some of their depleted comets have low C$_{2}$ but show C$_{3}$
within normal bounds.  This can be seen clearly in their Figure~10, comparing
panels a and b.

Using only a criterion for depleted comets that
$\log [Q(C_{2})/Q(CN)] \leq 0.02$, 
with no limit on the C$_{3}$, we add ten comets to our
list of depleted comets: 71P/Clark, 26P/Grigg-Skjellerup, 
75P/Kohoutek, C/1988~A1 (Liller), 31P/Schwassmann-Wachmann~2, C/1988~J1
(Shoemaker-Holt),
C/1986~V1 (Sorrells),
62P/Tsuchinshan~1, 
76P/West-Kohoutek-Ikemura, and C/1989~A1 (Yanaka).  With this definition,
we find 15/59 depleted comets, or 25\%.  This is consistent with the
quantity found by AH95.  These 10 comets all show ratios
for C$_{3}$ production at an average value or below.  However, none
are close to our defined value for C$_{3}$ depletion.

AH95 also noted that some comets were enhanced in C$_{2}$/CN. 
We too see
such a trend, although not necessarily the same comets.  We both
found C/1980~Y1 (Bradfield) to be enhanced in C$_{2}$.
In addition, 68P/Klemola has a strong C$_{2}$ enhancement.

F09 did not observe C$_{3}$ in their bandpass and so their
definition of depleted comets was based entirely on the C$_{2}$ abundance.
F09 observed a total of 92 comets, with 50 in their restricted group.
F09 broke their comets into more groups than just
typical and depleted.  However, all comets not in their typical group
had C$_{2}$ depleted relative to CN (except possibly C/1988~Y1 (Yanaka)).
They found 30\% depleted comets.

LS11 observed 26 comets.  They did not
define a subset of comets as part of a restricted group.  Many of
their comets were observed on only 1 night.  LS11
chose to derive new Haser model scale lengths for their data set
because they did not feel that any of the published scale lengths
fit their data well.  In addition, they derived a heliocentric 
distance scaling for those scale lengths that was different 
(not $R_h^2$) than 
most researchers use.  It was not clear if they used 
mean scale lengths or used different scale
lengths for each comet.  In addition, it appears from
their figures that they did not remove the underlying solar continuum
before computing the column densities, though they did fit
a continuum before integration.  For dusty comets, this will
increase the difficulty of detecting the gas.
Once they had computed their production rates in this manner, 
LS11
used the AH95 definition of depleted comets, 
finding all their Jupiter family comets depleted (including 4P/Faye where
they did
not detect any C$_{2}$ and 46P/Wirtanen that has a much higher C$_{2}$/CN
ratio in their data than this cutoff).   They also indicated that as a class
the long period comets were borderline depleted.  LS11 did not 
attempt to convert the AH95 limits to their 
scale lengths so it is not obvious that the comets they claim to be
depleted would be depleted if they derived a cut-off from their data
using their scale lengths.  Using the AH95 definition
and the data from their Table~11, there are 13/26 comets that are
depleted, or 50\% (this includes several with no C$_{2}$ detection
that they take to indicate a comet is depleted).

Table~\ref{summary} lists the 110 comets we observed that
had at least CN detected, along with a summary of our findings. 
%\begin{landscape}
\begin{table*}[ht!]
\scriptsize
\caption{Summary of Abundance Patterns}\label{summary}
\vspace*{5pt}
\begin{tabular}{lccccc}
\hline
\multicolumn{1}{c}{Comet} & Dynamical & This &A'Hearn & Fink & Langlund-Shula \\
  & Type & Paper & {\it et al.}. &        & \& Smith \\
  &      &       & (1995) & (2009) & (2011) \\
\hline
{\bf 49P/Arend-Rigaux } & JFC & high NH, C$_{3}$& Typical &  &   \\
{\bf C/1982 M1 (Austin) } & LPC &  Typical & Enhanced & & \\
{\bf C/1984 N1 (Austin) } & LPC & Typical & Typical & & \\
{\bf C/1989 X1 (Austin) } & LPC & low NH & Typical & Typical& \\
{\bf 19P/Borrelly } & JFC & \underline{Depleted} + low CH & Depleted & low C$_{2}$ & \\
\hspace{1em}140P/Bowell-Skiff  & JFC & low C$_{3}$, high CH & & & \\
\hspace{1em}C/1979 Y1 (Bradfield)  & LPC & \underline{Depleted} & Typical & & \\
{\bf C/1980 Y1 (Bradfield) } & LPC & high CH, C$_{2}$; low NH$_{2}$, C$_{3}$ & Enhanced & & \\
{\bf C/1987 P1 (Bradfield)  } & LPC &Typical & Typical & & \\
\hspace{1em}16P/Brooks 2  & JFC & Limits & Depleted & & \\
{\bf 23P/Brorsen-Metcalf } & HTC &  high OH, CH; low NH, NH$_{2}$ & Typical & Typical & \\
\hspace{1em}C/1983 O1 (Cernis)  & LPC & high C$_{3}$ & Depleted \\
\hspace{1em}C/1990 E1 (CKN)  & LPC & low CH & & \\
\hspace{1em}C/1980 O1 (CP)  & LPC & Limits & Typical & & \\
{\bf 67P/CG } & JFC & Typical & Depleted & low C$_{2}$ & \\
{\bf 71P/Clark } & JFC & low C$_{2}$ (Depleted) & low C$_{2}$ & & \\
{\bf 32P/Comas Sola } & JFC & \underline{Depleted} & &\\
{\bf 27P/Crommelin } & HTC & low NH$_{2}$ & Typical & & \\
{\bf 6P/d'Arrest } & JFC & Typical & Typical & Typical & \\
\hspace{1em}122P/de Vico  & HTC & low C$_{3}$ & Typical & \\
\hspace{1em}79P/dTH  & JFC & Limits & & & \\
{\bf 2P/Encke } & JFC & low NH, high CH & Typical & Typical & \\
\hspace{1em}4P/Faye  & JFC & Limits & Depleted & & Depleted \\
\hspace{1em}37P/Forbes  & JFC & Limits & & & \\
%\hline
%\end{tabular}
%\end{table}
%\newpage
%\begin{table}
%\begin{tabular}{lccccc}
%\multicolumn{6}{c}{Table~\ref{summary} continued} \\
%\hline
%\multicolumn{1}{c}{Comet} & Dynamical & This &A'Hearn & Fink & Langlund-Shula \\
%  & Type & Paper & {\it et al.}. &        & \& Smith \\
%  &      &       & (1995) & (2009) & (2011) \\
%\hline
\hspace{1em}78P/Gehrels 2  & JFC & Limits & Typical & & \\
{\bf 21P/GZ } & JFC & \underline{Depleted} & Depleted & low C$_{2}$, NH$_{2}$ & \\
\hspace{1em}84P/Giclas  & JFC & Limits & & & \\
{\bf 26P/GS } & JFC & low C$_{2}$, NH$_{2}$ (Depleted) & Depleted & & \\
\hspace{1em}65P/Gunn  & JFC & Limits & Depleted & & \\
{\bf C/1995 O1 (Hale-Bopp) } & LPC & high C$_{3}$, CH & & Typical & Depleted \\
{\bf 1P/Halley } & HTC & high CH, C$_{2}$ & Typical & Typical \\
\hspace{1em}51P/Harrington  & JFC & Typical & \\
\hspace{1em}52P/Harrington-Abell  & JFC & Limits & & \\
\hspace{1em}C/1995 Q2 (HD)  & LPC & Limits & & \\
{\bf 1985 R1 (Hartley-Good) } & LPC & high NH$_{2}$ & Typical &  Typical& \\
\hspace{1em}161P/Hartley-IRAS  & JFC & high NH$_{2}$ & Typical & & \\
\hspace{1em}111P/HRC  & JFC & Limits & & & \\
\hspace{1em}45P/HMP  & JFC & low C$_{3}$ & Typical & low C$_{2}$ & \\
{\bf 88P/Howell } & JFC & high C$_{3}$ & Typical & & \\
\hspace{1em}C/1995 Y1 (Hyakutake)   & LPC & low OH, NH & & Typical & \\
{\bf C/1996 B2 (Hyakutake)  } & LPC & Typical & & Typical & \\
{\bf 153P/Ikeya-Zhang } & LPC & low OH, C$_{3}$; high CH & & Typical \\
{\bf C/1983 H1 (IAA) } & HTC & Typical & Typical & & \\
\hspace{1em}126P/IRAS  & JFC & Limits & Depleted & & \\
\hspace{1em}C/1983 O2 (IRAS) & LPC & Limits & Typical & & \\
{\bf 58P/Jackson-Neujmin } & JFC & high NH, C$_{3}$ & & Typical & \\
{\bf 59P/Kearns-Kwee } & JFC & Typical & Typical & \\
{\bf 68P/Klemola } & JFC & high C$_{2}$ & Depleted & \\
\hline
\end{tabular}
\end{table*}
%\newpage
\begin{table*}[ht!]
\scriptsize
\vspace*{5pt}
\begin{tabular}{lccccc}
\multicolumn{6}{c}{Table~\ref{summary} continued} \\
\hline
\multicolumn{1}{c}{Comet} & Dynamical & This &A'Hearn & Fink & Langlund-Shula \\
  & Type & Paper & {\it et al.}. &        & \& Smith \\
  &      &       & (1995) & (2009) & (2011) \\
\hline
{\bf 75P/Kohoutek } & JFC & low C$_{2}$, NH$_{2}$ (Depleted) & & & \\
{\bf 22P/Kopff } & JFC & low OH & Typical & Typical & Depleted \\
{\bf 144P/Kushida } & JFC & Typical & & & \\
{\bf C/1987 A1 (Levy) } & LPC & low CH & & & \\
{\bf C/1990 K1 (Levy) } & LPC & high OH, C$_{3}$, C$_{2}$ & Typical & Typical & \\
{\bf C/1984 V1 (LR) } & LPC & Typical & Typical & & \\
{\bf C/1988 A1 (Liller) } & LPC & low C$_{2}$ (Depleted) & Typical & & \\
\hspace{1em}C/2001 A2 (LINEAR)  & LPC & Typical & & \\
{\bf C/2000 WM1 (LINEAR) } & LPC & \underline{Depleted} + low CH & & & \\
\hspace{1em}C/2006 VZ13 (LINEAR)  & LPC & low OH & & & \\
\hspace{1em}93P/Lovas 1  & JFC & high C$_{3}$ & & Typical & Depleted \\
\hspace{1em}C/2007 E2 (Lovejoy)  & LPC & low OH & & & Typical \\
{\bf C/1994 T1 (Machholz) } & LPC & low NH & & low C$_{2}$ & \\
{\bf C/2004 Q2 (Machholz) } & LPC & Typical & & & \\
\hspace{1em}141P/Machholz 2-A  & JFC & high CH & & Typical & \\
\hspace{1em}115P/Maury  & JFC & Limits & & & \\
{\bf C/1993 Y1 (MR) } & LPC & low CH & & Typical & \\
{\bf C/1980 V1 (Meier) } & LPC & low NH & Typical & & \\
\hspace{1em}C/1993 A1 (Mueller)  & LPC & low OH, high C$_{3}$ & & \\
{\bf 28P/Neujmin 1 } & JFC & high C$_{3}$ & Typical & & \\
{\bf C/1987 B1 (NTT) } & LPC & Typical & Typical & & \\
\hspace{1em}C/1992 W1 (Oshita)  & LPC & limits & & & \\
{\bf C/1980 Y2 (Panther) } & LPC & low C$_{3}$ & Typical & & \\
{\bf 80P/Peters-Hartley } & JFC & low CH & & Typical & \\
%\hline
%\end{tabular}
%\end{table}
%\newpage
%\begin{table}
%\begin{tabular}{lccccc}
%\multicolumn{6}{c}{Table~\ref{summary} continued} \\
%\hline
%\multicolumn{1}{c}{Comet} & Dynamical & This &A'Hearn & Fink & Langlund-Shula \\
%  & Type & Paper & {\it et al.}. &        & \& Smith \\
%  &      &       & (1995) & (2009) & (2011) \\
%\hline
\hspace{1em}83P/Russell 1  & JFC & limits & & & \\
\hspace{1em}91P/Russell 3  & JFC & limits & & & \\
{\bf 24P/Schaumasse } & JFC & Typical & & Typical & \\
\hspace{1em}106P/Schuster  & JFC & limits & & \\
{\bf 31P/SW2 } & JFC & low C$_{2}$ (Depleted) & Depleted & & \\
{\bf 102P/Shoemaker 1 } & JFC & high CH, NH$_{2}$ & & & \\
\hspace{1em}C/1984 K1 (Shoemaker)  & LPC & high C$_{3}$ & Enhanced & & \\
\hspace{1em}C/1984 U2 (Shoemaker)  & LPC & limits/\underline{Depleted} & Depleted & & \\
\hspace{1em}C/1987 H1 (Shoemaker)  & LPC & limits & & & \\
\hspace{1em}C/1989 A6 (Shoemaker)  & LPC & high C$_{3}$, low C$_{2}$ (Depleted) & & & \\
{\bf C/1988 J1 (SH) } & LPC & low C$_{2}$ (Depleted) & & & \\
\hspace{1em}192P/Shoemaker-Levy 1   & JFC & limits & & & \\
{\bf C/2002 E2 (SM) } & LPC & low NH &  & & \\
{\bf C/1986 V1 (Sorrells) } & LPC & low C$_{2}$, NH$_{2}$ (Depleted) & Depleted & & \\
{\bf 38P/Stephan-Oterma } & HTC & high CH & Typical & & \\
\hspace{1em}C/1983 J1 (SSF)  & LPC & low C$_{3}$ & Typical & & \\
{\bf 64P/Swift-Gehrels } & JFC & Typical & & & \\
\hspace{1em}109P/Swift-Tuttle  & HTC & low NH, CH & Typical & Typical & \\
{\bf C/1996 B1 (Szczepanski) } & LPC & Typical & & Typical & \\
\hspace{1em}98P/Takamizawa  & JFC & Limits & Depleted & & \\
\hspace{1em}C/1994 J2 (Takamizawa)  & LPC & Typical & & low C$_{2}$ & \\
{\bf C/1994 G1-A (TL) } & LPC & Typical & & & \\
\hspace{1em}69P/Taylor  & JFC & limits & & & \\
{\bf 9P/Tempel 1 } & JFC & Typical & Typical & low C$_{2}$ & Depleted \\
\hline
\end{tabular}
\end{table*}
%\newpage
\begin{table*}[ht!]
\scriptsize
\begin{tabular}{lccccc}
\multicolumn{6}{c}{Table~\ref{summary} continued} \\
\hline
\multicolumn{1}{c}{Comet} & Dynamical & This &A'Hearn & Fink & Langlund-Shula \\
  & Type & Paper & {\it et al.}. &        & \& Smith \\
  &      &       & (1995) & (2009) & (2011) \\
\hline
{\bf 10P/Tempel 2 } & JFC & Typical & Typical & Typical & Depleted \\
\hspace{1em}55P/Tempel-Tuttle & HTC & high CH & & Typical & \\
\hspace{1em}C/1987 B2 (Terasako)  & LPC & low CH, C$_{2}$ (Depleted) & & \\
\hspace{1em}C1985 T1 (Thiele)  & LPC & high NH$_{2}$ & Typical & & \\
{\bf 62P/Tsuchinshan 1 } & JFC & low C$_{2}$ (Depleted) & Typical & Typical & \\
{\bf 8P/Tuttle } & HTC & low NH & Typical & & Typical \\
\hspace{1em}41P/TGK  & JFC & high C$_{3}$ & & & \\
\hspace{1em}40P/Vaisala  & JFC & limits & & & \\
{\bf 76P/WKI } & JFC & high NH, low C$_{2}$ (Depleted) & & & \\
{\bf 81P/Wild 2 } & JFC & \underline{Depleted} & Depleted & low C$_{2}$ & Depleted \\
{\bf C/1986 P1 (Wilson) } & LPC & low NH$_{2}$ & & & \\
\hspace{1em}114P/Wiseman-Skiff  & JFC & limits & & \\
\hspace{1em}43P/Wolf-Harrington  & JFC & Limits/\underline{Depleted} & Depleted & low C$_{2}$, NH$_{2}$ & \\
{\bf C/1989 A1 (Yanaka) } & LPC & low C$_{2}$ (Depleted) & & Typical & \\
\hline
\end{tabular}

\end{table*}
%\end{landscape}
As before, the restricted comets are listed in bold and the remainder
are indented.  The group of comets that are depleted under our
most restrictive definition of low C$_{2}$ \underline{and} C$_{3}$ are
noted as ``Depleted" (with the word depleted underlined). 
The additional 10 comets
with low C$_{2}$ but not low C$_{3}$ are listed as ``(depleted)".  Comets
that have all of their production rate ratios within the normal ranges
shown for the restricted set in Table~\ref{ratios}
are marked ``Typical".  There are comets that are not depleted but have
at least one production rate ratio outside of the normal ranges.  We note the
species that are either high or low.  Inspection of the table
shows that some comets have only one species outside a normal
range while others (e.g. C/1990~Y1 (Bradfield)) have many species 
outside the normal range.  Comets with only limits are so noted.

Included in Table~\ref{summary} are the findings of the other studies
discussed above for comets observed in common.  When comparing our
comets with the findings of others, it is important to remember
that only C$_{2}$ was used in other studies as a criterion for being
called depleted,
so that any comet where we saw normal C$_{2}$ but other species out
of range would still be typical in other studies.

Inspection of Table~\ref{summary} shows that we are generally in
(remarkably good)
agreement on how we classify comets, with some exceptions.  
As noted above when we examined scale lengths, placement of the cut-off
for defining depletion has an important affect on our classifications.
In the discussion here, keep in mind our reservations about the
definition of depletion used by LS11.
In our restricted data set, we find C/1982~M1 (Austin)
typical while AH95 find it enhanced.  F09 and 
AH95 find 67P/Churyumov-Gerasimenko depleted but
we find it typical.  C/1995~O1 (Hale-Bopp) is typical in our data
and that of F09 but found to be depleted in LS11.
AH95 find 68P/Klemola to be depleted but we find
it to be {\it enhanced} in C$_{2}$.  22P/Kopff and
10P/Tempel~2 are typical for us,
F09 and AH95, but LS11 find them depleted.
9P/Tempel~1 appears normal to us and AH95, but depleted
to F09 and LS11.  C/1988~A1 (Liller), 62P/Tsuchinshan~1,
and C/1989~A1 (Yanaka) are depleted in our extended definition (Liller
and Tsuchinshan 1 barely so) but typical according to AH95
and/or F09.  C/1994~T1 (Machholz) is depleted for F09
but typical for us.
In addition, there are five comets not in our restricted set where
we see some differences with others.

If all of the depleted comets started in the same formation reservoir,
we would expect to see all of the depleted comets having the same
dynamical type.  Of course, the dynamical type does not necessarily
correspond to a single reservoir since there have been exchanges of
objects from one reservoir to another via gravitational perturbations.
For our 59 comets in the restricted data set, 27 are JFCs, 5 are
HTCs and 27 are LPCs.  Using our strict definition of carbon-chain
depleted comets, we find 4/5 of them are JFCs (80\%)
and 1 is an LPC (20\%).  Using the more relaxed definition requiring
only C$_{2}$ to be depleted, we find 10 of the depleted comets
are JFCs (67\%) and 5 are LPCs (33\%).  For a particular
dynamical type and the relaxed definition, 10/27 or 37\% of the
JFCs are depleted; 5/27 or 18.5\% of the LPCs are depleted.
We see no depleted HTCs, but with only 5 HTCs in our sample this might
not be significant.

In Table~\ref{dynamical}, we show the average production rate ratios as
a function of dynamical type,  combining depleted and typcial comets
in each dynamical type.
\begin{table*}[ht!]
\footnotesize
\caption{Production Rate Ratios as a function of Dynamical Type}\label{dynamical}
\vspace*{7pt}
\begin{tabular}{lcccccc}   % 7 cols
\hline
 & \multicolumn{6}{c}{Log Production Rate Ratio with respect to CN} \\
\cline{2-7}
 & OH & NH & C$_{3}$ & CH & C$_{2}$ & NH$_{2}$  \\
 &    &    &         &    & (average) &         \\
\hline
Jupiter Family & $2.15 \pm 0.32$ & $0.60 \pm 0.24$ & $-0.69 \pm 0.25$ &
  $0.11 \pm 0.19$ & $0.04 \pm 0.31$ & $-0.03 \pm 0.28 $ \\
\hspace*{1em}(27 comets) & (11 comets) & (12 comets) & (27 comets) & (14 comets) & (27 comets)& (11 comets)   \\ [5pt]
Halley Type & $2.76 $  & $0.25 $ & $-0.67 \pm 0.08$ &
  $0.46 \pm 0.28$ & $0.16 \pm 0.13$ & $0.10 \pm 0.34 $  \\
\hspace*{1em}(5 comets) & (1 comet) & (1 comet) & (5 comets) & (4 comets) & (5 comets)& (4 comets) \\ [5pt]
Long Period & $2.17 \pm 0.38$ & $0.36 \pm 0.18$ & $-0.68 \pm 0.12$ &
  $0.27 \pm 0.17$ & $0.18 \pm 0.11$ & $-0.17 \pm 0.22 $  \\
\hspace*{1em}(27 comets) & (11 comets) & (14 comets) & (27 comets) & (23 comets) & (27 comets)& (11 comets) \\ [5pt]
\hline
\end{tabular}
\end{table*}
Inspection of Table~\ref{dynamical}
shows that the averages are indistinguishable for OH (with only 1
Halley Type comet with OH the deviation of this group is not meaningful), 
or C$_{3}$.  CH and NH$_{2}$ might show differences with dynamical
type but the error
bars are large (and overlapping) and there is no obvious sequence with
current orbit size.   For NH, there appears to be a significant difference
between JFCs and LPCs (again HTCs do not have enough data to be
meaningful) but the error bars are overlapping.  For C$_{2}$, there
is a significant difference between JFCs and other dynamical
types, though LPCs and HTCs are identical.  This trend for C$_{2}$ is
consistent with us finding many more JFCs that are depleted
and is in agreement with the trend found by AH95.
We concur with their conclusion that the compositional similarities
between dynamical types
are suggestive that the interiors of comets are similar to their
exteriors.
In contrast, LS11 reported a trend that the C$_{2}$,
NH and NH$_{2}$ increased in a regular fashion as the comet orbits
became larger (though their Table~12 shows the C$_{2}$/CN decreasing
from HTCs to LPCs and being the same for dynamically new and HTCs).

Perhaps the differences we see for a large number of JFCs is
the result of evolutionary differences instead of formational differences.
While we cannot categorically refute this concept, we point to
observations of comet 73P/Schwassmann-Wachmann 3 (SW3) as evidence to the
contrary.  Comet SW3 split into several pieces in 1995.  It made
a close approach to the Earth in 2003 and was extensively studied.
73P/Schwassmann-Wachmann 3 showed no compositional differences
between different pieces (Kobayashi {\it et al.} 2007,
Jehin {\it et al.} 2008, Weaver {\it et al.} 2008, Schleicher and Bair 2011),
suggesting that the comet was the same composition throughout. 
SW3 is a JFC that is depleted.  Therefore, its depletion is not
just a surface effect.
\nocite{scba11,jeetalsw3,weetalsw3,koetalsw3}

\begin{figure}
\includegraphics[scale=0.33,angle=270]{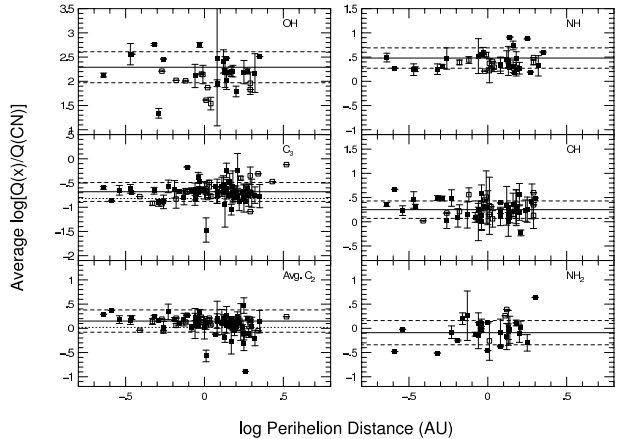}
\caption[fig15]{The production rate ratios are plotted
as a function of the perihelion distances of the current cometary orbits.
The symbols used are the same as in Figure~\ref{avgwlim}, as are the
definitions of the horizontal lines.  There are no believable trends
of the ratios as a function of perihelion distance.
}
\label{peritrend}
\end{figure}

AH95 found a trend of dust-to-gas ratio as a function of cometary
perihelion distance (their Figure~4).  With our small apertures, we do
not feel comfortable computing A$f\rho$ so cannot comment on whether
our data show such a trend. We did, however, examine the production rate
ratios as a function of cometary perihelion distance.  These are shown
in Figure~\ref{peritrend}.
We find no believable trend of the production rate ratios with
perihelion distance.  

\section{Summary}

When we began this survey in 1980, we fully expected that we would find
some comets with very different spectra than the vast majority of
comets.  Instead, we found a mostly homogeneous group of objects with
subtle compositional differences.  In particular, we found:

\begin{itemize}
\item The vast majority of comets have remarkably similar compositions.

\item There exists a group of comets which are depleted in the carbon-chain
molecules.  Using a very strict definition of carbon-chain depletion
requiring depletion in both C$_{2}$ and C$_{3}$, we find that 9\% of
the comets are depleted.  Using a more relaxed definition that requires
only the C$_{2}$ be depleted (similar to other authors), we
find 25\% of the comets are depleted.  This is consistent with
other authors.

\item For the most part, we agree with other studies on which
individual comets show depletions.

\item Carbon-chain depleted comets can be of any dynamical type (though
we did not observe any HTCs with depletion our HTC sample is very small).

\item Using the relaxed definition, two thirds of the depleted comets are JFCs
and one third are LPCs (with the stricter definition it is 80\% and
20\% respectively).

\item Depleted comets make up 37\% of the JFCs and 18.5\% of the LPCs.

\item On average, there are few differences in composition with
increasing current orbit semi-major axis.  However, C$_{2}$ and NH
may show significant trends  of lower C$_{2}$ and higher NH for JFCs than
for LPCs.

\end{itemize}

\vspace*{0.75in}
\begin{center}Acknowledgments\end{center}
This research was supported by NASA Grants NNG04G162G, NNX08A052G and
predecessor grants.  McDonald Observatory is operated by The University of
Texas at Austin.  We thank David Schleicher for performing a cluster
analysis for us that, unfortunately, did not clarify the groupings.
We thank Dr. Michael A'Hearn and Dr. Uwe Fink for very helpful
suggestions to clarify the paper and for identifying some errors.

%\newpage
%\bibliography{/home/barolo/comet/anita/documents/apj,/home/barolo/comet/anita/documents/master}

\begin{thebibliography}{}

\bibitem{adetal03}
{\sc \'{A}d\'{a}mkovics, M., G.~A.  Blake, and B.~J.  McCall}  2003.
  Observations of rotationally resolve {C$_{3}$} in translucent sight lines.
\newblock {\it Ap. J. \/} {\bf 595}, 235--246.

\bibitem{ahbifemi85}
{\sc A'Hearn, M.~F., P.~V.  Birch, P.~ Feldman, and R.~L.  Millis}  1985. Comet
  {E}ncke: {G}as production and lightcurve.
\newblock {\it Icarus \/} {\bf 64}, 1--10.

\bibitem{ahmi80}
{\sc A'Hearn, M.~F. and R.~L.  Millis}  1980. Abundance correlations among
  comets.
\newblock {\it A. J. \/} {\bf 85}, 1528--1537.

\bibitem{ahetal95}
{\sc A'Hearn, M.~F., R.~L.  Millis, D.~G.  Schleicher, D.~J.  Osip, and P.~V.
  Birch}  1995. The ensemble properties of comets: Results from narrowband
  photometry of 85 comets, 1976--1992.
\newblock {\it Icarus \/} {\bf 118}, 223--270.

\bibitem{argrpe69}
{\sc Arvesen, J.~C., R.~J.  Griffin, {Jr.}, and B.~D.  Pearson, {Jr.}}  1969.
  Determination of extraterrestrial {S}olar spectral irradiance from a research
  aircraft.
\newblock {\it Appl. Opt. \/} {\bf 8}, 2215--2232.

\bibitem{clla82}
{\sc Clegg, R. E.~S. and D.~L.  Lambert}  1982. On c$_{\hbox{3}}$ molecules in
  diffuse interstellar clouds.
\newblock {\it M.N.R.A.S. \/} {\bf 201}, 723--733.

\bibitem{co86haser}
{\sc Cochran, A.~L.}  1986. A reevaluation of the {H}aser model scale lengths
  in comets.
\newblock {\it A. J. \/} {\bf 90}, 2609--2614.

\bibitem{co87corr}
{\sc Cochran, A.~L.}  1987. Another look at abundance correlations among
  comets.
\newblock {\it A. J. \/} {\bf 93}, 231--238.

\bibitem{coba85}
{\sc Cochran, A.~L. and E.~S.  Barker}  1985. Spatially resolved
  spectrophotometry of comet {P}/{Stephan-Oterma}.
\newblock {\it Icarus \/} {\bf 62}, 72--81.

\bibitem{coba87gz}
{\sc Cochran, A.~L. and E.~S.  Barker}  1987. Comet {G}iacobini-{Z}inner: {A}
  normal comet?
\newblock {\it A. J. \/} {\bf 93}, 239--243.

\bibitem{cobacari09}
{\sc Cochran, A.~L., E.~S.  Barker, M.~D.  Caballero, and J.~ G{y\"o}rgey-Ries}
   2009. Placing the {D}eep {I}mpact mission into context: Two decades of
  observations of {9P/T}empel~1 from {McD}onald {O}bservatory.
\newblock {\it Icarus \/} {\bf 199}, 119--128.

\bibitem{cobaco80}
{\sc Cochran, A.~L., E.~S.  Barker, and W.~D.  Cochran}  1980.
  Spectrophotometric observations of {P}/{Schwassmann-Wachmann} 1 during
  outburst.
\newblock {\it A. J. \/} {\bf 85}, 474--477.

\bibitem{cobarast92}
{\sc Cochran, A.~L., E.~S.  Barker, T.~F.  Ramseyer, and A.~D.  Storrs}  1992.
  The {M}c{D}onald {O}bservatory {F}aint {C}omet {S}urvey: {G}as production in
  17 comets.
\newblock {\it Icarus \/} {\bf 98}, 151--162.

\bibitem{coco90}
{\sc Cochran, A.~L. and W.~D.  Cochran}  1990.
\newblock Observations of {CH} in comets {P/B}rorsen-{M}etcalf and {P/H}alley.
\newblock In {\em Workshop on Observations of Recent Comets (1990)} (W.~F.
  Huebner, P.~A.  Wehinger, J.~ Rahe, and I.~ Konno, Eds.) pp.~22--27 Southwest
  Research Institute San Antonio, TX.

\bibitem{coco91sw1p1}
{\sc Cochran, A.~L. and W.~D.  Cochran}  1991. The first detection of {CN} and
  the distribution of {CO$^+$} gas in the coma of comet
  {P/Schwassmann-Wachmann} 1.
\newblock {\it Icarus \/} {\bf 90}, 172--175.

\bibitem{cocoba82}
{\sc Cochran, A.~L., W.~D.  Cochran, and E.~S.  Barker}  1982.
  Spectrophotometry of comet {S}chwassmann-{W}achmann 1. {II}. {I}ts color and
  {CO}$^{\hbox{+}}$ emission.
\newblock {\it Ap. J. \/} {\bf 254}, 816--822.

\bibitem{cocobast91}
{\sc Cochran, A.~L., W.~D.  Cochran, E.~S.  Barker, and A.~D.  Storrs}  1991.
  The development of the {CO$^+$} coma of comet {P/Schwassmann-Wachmann} 1.
\newblock {\it Icarus \/} {\bf 92}, 178--183.

\bibitem{cogrba89}
{\sc Cochran, A.~L., J.~R.  Green, and E.~S.  Barker}  1989. Are low activity
  comets different from more active comets?
\newblock {\it Icarus \/} {\bf 79}, 125--144 plus erratum {\it Icarus} {\bf
  80}, 446, 1989.

\bibitem{cosc93}
{\sc Cochran, A.~L. and D.~G.  Schleicher}  1993. Observational constraints on
  the lifetime of cometary {H$_{2}$O}.
\newblock {\it Icarus \/} {\bf 105}, 235--253.

\bibitem{cosl994}
{\sc Cochran, A.~L., A.~L.  Whipple, P.~J.  MacQueen, P.~J.  Shelus, R.~W.
  Whited, and C.~F.  Claver}  1994. Pre-impact characterization of comet
  {Shoemaker-Levy 9}.
\newblock {\it Icarus \/} {\bf 112}, 528--532.

\bibitem{cowh93}
{\sc Cochran, A.~L. and B.~D.  White}  1993. Spectroscopic observations of
  {CO$^+$} in {P/S}chwassmann-{W}achmann 1: 1978-1991.
\newblock In {\em Proceedings of the Lenggries Workshop on the Activity of
  Distant Comets} (W.~F.  Huebner, H.~U.  Keller, D.~ Jewitt, J.~ Klinger, and
  R.~ West, Eds.)
\newblock pp.~29--38 Southwest Research Institute, San Antonio, TX.

\bibitem{cofi93}
{\sc Combi, M.~R. and U.~ Fink}  1993. {P/Halley}: {E}ffects of time-dependent
  production rates on spatial emission profiles.
\newblock {\it Ap. J. \/} {\bf 409}, 790--797.

\bibitem{fi94}
{\sc Fink, U.}  1994. The trend of production rates with heliocentric distance
  for comet {P/Halley}.
\newblock {\it Ap. J. \/} {\bf 423}, 461--473.

\bibitem{fi09}
{\sc Fink, U.}  2009. A taxonomic survey of comet composition 1985 -- 2004
  using {CCD} spectroscopy.
\newblock {\it Icarus \/} {\bf 201}, 311--334.

\bibitem{fihi96}
{\sc Fink, U. and M.~D.  Hicks}  1996. A survey of 39 comets using {CCD}
  spectroscopy.
\newblock {\it Ap. J. \/} {\bf 459}, 720--743.

\bibitem{ha57}
{\sc Haser, L.}  1957.
\newblock Distribution d'intensit\'{e} dans la t\^{e}t\^{e} d'une com\`{e}te.
\newblock Liege Inst. Astrophysics Reprint
\newblock No. 394.

\bibitem{ihpidoco11}
{\sc Ihalawela, C.~A., D.~M.  Pierce, G.~R.  Dorman, and A.~L.  Cochran}  2011.
  The spatial distribution of {OH} and {CN} radicals in the coma of comet
  {E}ncke.
\newblock {\it Ap. J. \/} {\bf 741}, 89.

\bibitem{jeetalsw3}
{\sc {Jehin}, E., J.~ {Manfroid}, H.~ {Kawakita}, D.~ {Hutsem{\'e}kers}, M.~
  {Weiler}, C.~ {Arpigny}, A.~ {Cochran}, O.~ {Hainaut}, H.~ {Rauer}, R.~
  {Schulz}, and J.-M.  {Zucconi}}  2008.
\newblock Optical spectroscopy of the {B} and {C} fragments of comet
  {73P/Schwassmann-Wachmann 3} at the {ESO VLT}.
\newblock LPI Contributions {\bf 1405}, p.8319.

\bibitem{joalsi89}
{\sc J{\o}rgensen, U.~G., J.~ Alml{\"o}f, and P.~E.~M.  Siegbahn}  1989.
  Complete active space self-consistent field calculations of the vibrational
  band strengths for {C$_{3}$}.
\newblock {\it Ap. J. \/} {\bf 343}, 554--561.

\bibitem{kiahco89}
{\sc Kim, S.~J., M.~F.  A'Hearn, and W.~D.  Cochran}  1989. {NH} emissions in
  comets: {F}luorescence vs collisions.
\newblock {\it Icarus \/} {\bf 77}, 98--108.

\bibitem{koetalsw3}
{\sc {Kobayashi}, H., H.~ {Kawakita}, M.~J.  {Mumma}, B.~P.  {Bonev}, J.-i.
  {Watanabe}, and T.~ {Fuse}}  2007. Organic volatiles in comet
  {73P-B/Schwassmann-Wachmann 3} observed during its outburst: A clue to the
  formation region of the {J}upiter-{F}amily comets.
\newblock {\it Ap. J.(Letters) \/} {\bf 668}, L75--L78.

\bibitem{lasm11}
{\sc Langland-Shula, L.~E. and G.~H.  Smith}  2011. Comet classification with
  new methods for gas and dust spectroscopy.
\newblock {\it Icarus \/} {\bf 213}, 280--322.

\bibitem{nesp84}
{\sc Newburn, R.~L. and H.~ Spinrad}  1984. Spectrophotometry of 17 comets.
  {I}. {T}he emission features.
\newblock {\it A. J. \/} {\bf 89}, 289--309.

\bibitem{olhobr85}
{\sc Oliversen, R.~J., J.~M.  Hollis, and L.~W.  Brown}  1985. {C$_{2}$}
  imagery of the inner coma of comet {Iras-Araki-Alcock}.
\newblock {\it Icarus \/} {\bf 63}, 339--346.

\bibitem{raetal92}
{\sc Randall, C.~E., D.~G.  Schleicher, R.~G.  Ballou, and D.~J.  Osip}  1992.
  Observational constraints on molecular scalelengths and lifetimes in comets.
\newblock {\it Bull. AAS \/} {\bf 24}, 1002.

\bibitem{roetal02}
{\sc Roueff, E., P.~ Felenbok, J.~H.  Black, and C.~ Gry}  2002. Interstellar
  {C$_{3}$} toward {HD 210121}.
\newblock {\it Astr. and Ap. \/} {\bf 384}, 629--637.

\bibitem{sc10}
{\sc Schleicher, D.~G.}  2010. The fluorescence efficiencies of the {CN} violet
  bands in comets.
\newblock {\it A. J. \/} {\bf 140}, 973--984.

\bibitem{scah82}
{\sc Schleicher, D.~G. and M.~F.  A'Hearn}  1982. {OH} fluorescence in comets:
  {F}luorescence efficiency of the ultraviolet bands.
\newblock {\it Ap. J. \/} {\bf 258}, 864--877.

\bibitem{scba11}
{\sc Schleicher, D.~G. and A.~N.  Bair}  2011. The composition of the interior
  of comet {73P/Schwassmann-Wachmann 3}: Results from narrowband photometry of
  multiple components.
\newblock {\it A. J. \/} {\bf 141}, 177--.

\bibitem{st77}
{\sc Stone, R. P.~S.}  1977. Spectral energy distributions of standard stars of
  intermediate brightness. {II}.
\newblock {\it Ap. J. \/} {\bf 218}, 767--769.

\bibitem{tagi77}
{\sc Tatum, J.~B. and M.~J.  Gillespie}  1977. The cyanogen abundance of
  comets.
\newblock {\it Ap. J. \/} {\bf 218}, 569--572.

\bibitem{tewy89}
{\sc Tegler, S. and S.~ Wyckoff}  1989. {NH}$_2$ fluorescence efficiencies and
  the {NH$_{3}$} abundance in comet {H}alley.
\newblock {\it Ap. J. \/} {\bf 343}, 445--449.

\bibitem{weetalsw3}
{\sc {Weaver}, H.~A., C.~M.  {Lisse}, M.~ {Mutchler}, P.~L.  {Lamy}, I.~
  {Toth}, W.~T.  {Reach}, and J.~ {Vaubaillon}}  2008.
\newblock Hubble investigation of the {B} and {G} fragments of comet
  {73P/Schwassmann-Wachmann 3}.
\newblock LPI Contributions, {\bf 1405}, p. 8248.

\bibitem{wi67}
{\sc Wing, R.~F.}  1967.
\newblock {\em Infrared Spectrophotometry of Red Giant Stars}.
\newblock PhD thesis U. C. Berkeley.

\end{thebibliography}
%\bibliographystyle{icarus}

\end{document}